\documentclass[twocolumn,preprintnumbers]{revtex4}

\usepackage[dvipdfmx]{hyperref}
\usepackage{graphicx}
\usepackage{epsf}
\usepackage{amsmath,amssymb}
\usepackage{color}

\def\Pb#1#2#3{{}^{#1#2#3}\textrm{Pb}}
\def\Ni#1#2{{}^{#1#2}\textrm{Ni}}
\def\Ca#1#2{{}^{#1#2}\textrm{Ca}}
\def\Sn#1#2#3{{}^{#1#2#3}\textrm{Sn}}
\def\SnA{{}^{A}\textrm{Sn}}

\usepackage{ulem}

\begin{document}
\preprint{KUNS-2877}
\title{Shell effect in $A=$116--124 Tin isotopes  investigated 
using isotopic analysis of proton scattering at 295~MeV}

\author{Yoshiko Kanada-En'yo}
\affiliation{Department of Physics, Kyoto University, Kyoto 606-8502, Japan}

\begin{abstract}
 Proton elastic scattering off Sn isotopes at $E_p=295$~MeV in the mass number range of $A=$116--124
was investigated using calculation employing relativistic impulse approximation~(RIA) with
theoretical densities obtained for the Sn isotopes from  
relativistic Hartree-Bogoliubov~(RHB) and nonrelativistic Skyrme Hartree-Fock-Bogoliubov~(SHFB)
calculations of spherical nuclei. 
In the RIA calculations, a modified version of the Murdock and Horowitz model that includes a density dependence in the effective nucleon-nucleon~($NN$) interaction was used.
A calculation 
using the theoretical density obtained from a relativistic calculation employing the DD-ME2 interaction
successfully reproduced the experimental data for $\Sn122(p,p)$, but it overestimated the  $\Sn116(p,p)$
and  $\Sn118(p,p)$ cross sections at backward angles. 
I found 
the normalization of the  experimental $\Sn120(p,p)$ and $\Sn124(p,p)$ cross sections to be inconsistent with 
the data for the other Sn isotopes, suggesting that it should be corrected. 
Isotopic analyses of the $(p,p)$ reactions combined with nuclear structure properties
were performed based on reaction calculations that used a model density modified from the DD-ME2 density
to optimize the neutron density of the Sn isotopes by fitting the isotopic cross section ratios.
The resulting optimized  density reproduced the experimental $(p,p)$ data for the series of Sn isotopes
from $\Sn116$ to $\Sn124$.
The neutron root-mean-square~(rms) radii and the skin thickness of the Sn isotopes obtained in the present analysis exhibited smooth  $A$ dependences in the range of $A=$116--124, which are  consistent 
with the theoretical predictions obtained using the DD-ME2 interaction but seem to contradict 
the experimental results determined from the $(p,p)$ data.
In a detailed analysis of the surface neutron density probed by proton elastic scattering, 
a signal of the shell effect at $N=66$ in Sn isotopes was found. 
\end{abstract}

\maketitle

\section{Introduction}\label{sec:introxution}
The neutron skin thickness $\Delta r_{np}$ in $N>Z$ nuclei
has recently emerged as an issue in understanding the neutron-matter equation of state
involving symmetry energy parameters that are connected with $\Delta r_{np}$ 
through predictions of theoretical structure 
calculations~\cite{RocaMaza:2011pm,Roca-Maza:2018ujj,Tsang:2012se}. 
Various experiments have been performed to determine $\Delta r_{np}$. 
For example, experimental values of $\Delta r_{np}$ in Sn and Pb isotopes 
have been obtained by measuring 
proton elastic scattering \cite{Ray:1979qv,Terashima:2008zza,Starodubsky:1994xt,Zenihiro:2010zz}, 
$X$-rays from antiprotonic atoms \cite{Trzcinska:2001sy,Klos:2007is}, parity-violating electron
scattering \cite{Abrahamyan:2012gp}, pionic probes~\cite{Friedman:2012pa,Tarbert:2013jze}, 
spin-dipole resonances measured with the $(^3\textrm{He},t)$ charge-exchange reaction~\cite{Krasznahorkay:1999zz}, 
and electric dipole polarization measured with 
polarized-proton inelastic scattering \cite{Tamii:2011pv,Piekarewicz:2012pp}.

The proton elastic scattering is a useful tool not only for determining the 
neutron skin thickness of an atomic nucleus but also for probing the  
density profile of the nucleus in detail---in particular, the surface neutron 
density---as has been done for various nuclei. For example, the experimental 
cross sections and analyzing powers measured with $(p,p)$ reactions 
at $E_p=800$~MeV~\cite{Ray:1978ws,Ray:1979qv,Hoffmann:1980kg}, 650~MeV~\cite{Ray:1978ws,Ray:1979qv,Hoffmann:1980kg}, and 295~MeV~\cite{Terashima:2008zza,Zenihiro:2010zz,Zenihiro:2018rmz}
have been utilized to extract the neutron density via reaction analyses. 
For the analyses of proton elastic scattering in the energy range of $E_p=100$--400~MeV,
Murdock and Horowitz proposed 
a reaction model based on the relativistic impulse approximation (RIA), with a meson-exchange model of 
effective nucleon-nucleon~($NN$)
interactions~(the MH model)~\cite{Horowitz:1985tw,Murdock:1986fs,RIAcode:1991}.
The original MH model was tuned by globally fitting the $(p,p)$ data, and 
the computational code for RIA+MH calculation has been widely used for the analyses of $(p,p)$ reactions.
Later, Sakaguchi and his collaborators proposed 
a modified version of the MH model that includes a density dependence of 
the effective $NN$ interaction~\cite{Sakaguchi:1998zz,Terashima:2008zza,Zenihiro:2010zz}; 
this is called the ddMH model in this paper.
The ddMH model was calibrated with  experimental $\Ni58(p,p)$ data at 295~MeV
for scattering angles $\theta_\textrm{c.m.}\lesssim 50^\circ$, and it 
successfully described the 295~MeV $(p,p)$ reactions of various target nuclei, including 
Sn~\cite{Terashima:2008zza}, Pb~\cite{Zenihiro:2010zz}, and Ca~\cite{Zenihiro:2018rmz} isotopes.

In a previous paper~\cite{Kanada-Enyo:2021}, I proposed a new method of reaction analysis that  combines
the proton elastic scattering 
with the isotopic systematics of nuclear structure for a series of isotopes. 
The method has been applied to the analysis of $\Pb204(p,p)$, $\Pb206(p,p)$, and $\Pb208(p,p)$  at 295~MeV to
obtain improved neutron densities and the root-mean-square~(rms) radii of the Pb isotopes 
from the experimental data. 
It has proven to be a useful tool for extracting neutron densities and rms radii
from $(p,p)$ cross sections for a series of isotopes, with less 
model dependence. Moreover, 
it can be applied to $(p,p)$ reactions to determine the neutron skin thickness
of other series of isotopes. 
An advantage of the isotopic analysis is that 
systematic errors in the experimental data obtained using the same experimental setup
can be reduced.
  
In this work, the 295-MeV proton scattering off Sn isotopes in the range
$A=116$--124 is investigated, 
for which high-quality data have been obtained from the experiment
by Terashima {\it et al.}~\cite{Terashima:2008zza}.
The structure of the Sn isotopes is calculated by using both
 relativistic Hartree-Bogoliubov~(RHB) and nonrelativistic Skyrme Hartree-Fock-Bogoliubov~(SHFB)
calculations of spherical nuclei. Using theoretical densities, the Sn$(p,p)$ reactions are calculated 
with the RIA+ddMH model 
in the same way as done in the previous paper~\cite{Kanada-Enyo:2021}. 
By comparing the theoretical results with the experimental $(p,p)$ data, 
the isotopic systematics of the structure 
and reaction properties are investigated. The isotopic analysis is performed using a model density 
to obtain an optimized neutron density for the Sn isotopes that can reproduce 
the experimental $(p,p)$ data. Structural properties such as the surface neutron densities and 
the rms radii of the neutron densities in the Sn isotopes are also discussed, which suggest
the shell effect at $N=66$ in the Sn isotopes.

The paper is organized as follows. The structure and reaction calculations 
are explained in Sec.~\ref{sec:calculations}, and the results obtained using the theoretical densities 
are presented in Sec.~\ref{sec:results}.
In Sec.~\ref{sec:analysis}, 
the isotopic analysis is performed using a model density to obtain the optimized density, 
and the results obtained are discussed.
Finally, a summary is given in Sec.~\ref{sec:summary}.

\section{Calculations of nuclear structure and proton elastic scattering}\label{sec:calculations}
\subsection{Structure calculations}
Structure calculations for even-even Sn isotopes from $\Sn114$ to $\Sn124$ 
were performed by employing both RHB and SHFB
calculations of spherical nuclei 
using the computational DIRHB code~\cite{Niksic:2014dra} and HFBRAD code~\cite{Bennaceur:2005mx},
respectively.
The spherical assumption is reasonable because 
most mean-field calculations obtain the ground states in this range of mass numbers wit only weak or no
deformation of the Sn isotopes.
In the RHB calculations, the
DD-ME2~\cite{Lalazissis:2005de} and DD-PC1~\cite{Niksic:2008vp} interactions were used, 
which are simply denoted as me2 and pc1, respectively, in this paper. 
In the SHFB calculation,  the SKM*~\cite{Bartel:1982ed} 
interaction with a mixed-type pairing force was used. 
The SHFB calculations with the SLy4~\cite{Chabanat:1997un} interaction were also performed 
to check the interaction dependence, but the resulting densities of the Sn isotopes were
similar to those obtained using the SKM* interaction; therefore, only the SKM* result are presented
in this paper. 
Note that these structure models were tuned to fit 
the binding energies and rms charge radii globally over  a wide range of 
mass numbers extending from $\Ca40$ to $\Pb208$. 

\subsection{Calculations of proton elastic scattering reactions}

Sn$(p,p)$ reactions at $E_p=295$~MeV were calculated using the RIA+ddMH model, 
which is a modified version of the RIA+MH model proposed by Sakaguchi {\it et al.} \cite{Sakaguchi:1998zz}. 
In the RIA+MH and RIA+ddMH models, real and imaginary nucleon-nucleus potentials are constructed by folding 
the target density with effective $NN$ interactions of the meson-exchange model.
The effective $NN$ interaction in the original RIA+MH model
contains energy dependences in the meson masses and coupling constants, and it was tuned to 
fit proton elastic -scattering data globally over the energy range of $100~\textrm{MeV}\le E_p \le 400~\textrm{MeV}$.
In the RIA+ddMH model,  density-dependent $\sigma$- and $\omega$-meson masses and coupling constants 
were introduced into the original effective $NN$ interaction of 
the relativistic Love–Franey~(RLF) parametrization~\cite{Horowitz:1985tw,Murdock:1986fs}. 
The density dependence is considered to be ``a medium effect'' of the effective $NN$ interaction, 
which contains various many-body effects that occur in proton elastic scattering such as 
Pauli blocking, multistep processes, and the medium effects on meson properties. 
The parameterization of the density dependence of the RIA+ddMH model has been calibrated
to fit the $^{58}\textrm{Ni}(p,p)$ data at $295~$MeV, and it has been updated
in Refs.~\cite{Terashima:2008zza,Zenihiro:2010zz} from the original version~\cite{Sakaguchi:1998zz}.
In the present work, latest parametrization of the RIA+MH model determined in 
Ref.~\cite{Zenihiro:2010zz} was used, which afforded better reproduction of 
the $^{58}\textrm{Ni}(p,p)$ data in the range of $\theta_\textrm{c.m.}\lesssim 50^\circ$ 
than an earlier version~\cite{Terashima:2008zza} used for the analysis of Sn$(p,p)$ reactions.

The RIA+ddMH calculation was performed using the theoretical densities of Sn the isotopes.
In an additional case, the RIA+MH calculations with the RLF parametrization and the Pauli-blocking effect
were performed to check the model dependence in the reaction calculations.
In the reaction calculations, the proton-nucleus potentials are obtained by folding the vector 
and scalar densities of the target nuclei with the effective $NN$ interaction. 
The theoretical neutron $(\rho_n(r))$ and proton $(\rho_p(r))$ 
densities were used for the neutron and proton vector densities, whereas $0.96\rho_n(r)$ 
and $0.96\rho_p(r)$ of the theoretical densities 
were used for the neutron and proton scalar 
densities, respectively, consistently in the RHB and SHFB calculations.
This treatment is the same as that done in 
the experimental analyses of Refs.~\cite{Terashima:2008zza,Zenihiro:2010zz} and as  adopted in my previous paper. 
Note that 
this prescription for the scalar density is considered a type of local-density approximation for
the $\sigma$-meson exchange term of the effective $NN$ interaction. 

\section{Results}\label{sec:results}

\subsection{Densities and radii of Sn}

\begin{figure}[!h]
\includegraphics[width=8 cm]{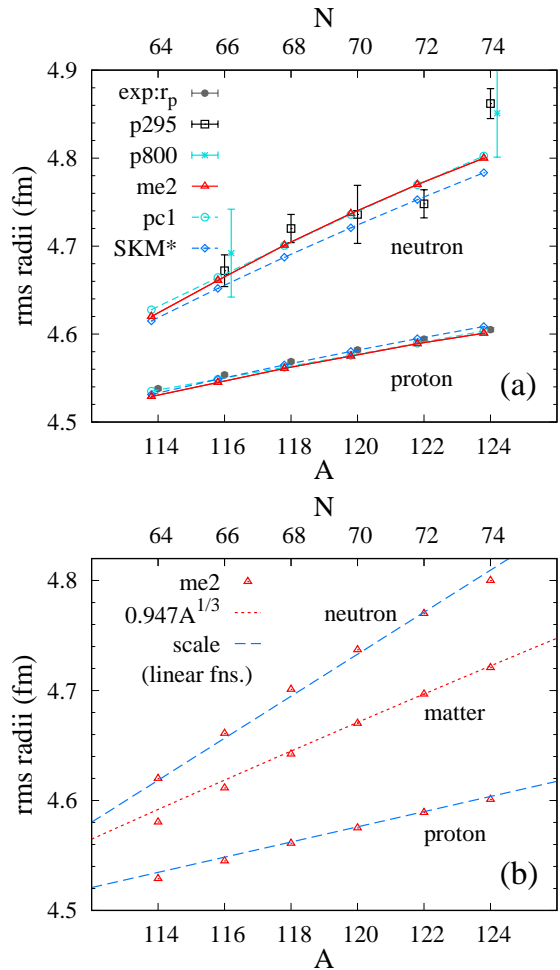}
  \caption{(a) Rms radii of the neutron and proton density distributions in the Sn isotopes. 
The theoretical values are the RHB (me2 and pc1) calculations and 
the SHFB (SKM*) calculation. 
The experimental data  are the $r_n$ values
obtained from the $(p,p)$ reactions at both 295~MeV\cite{Terashima:2008zza} and 800~MeV~\cite{Ray:1979qv},
and the $r_p$ values are obtained from experimental data for the rms charge radii from isotope-shift measurements~\cite{Angeli:2013epw}. 
(b) The rms radii of the matter, proton, and neutron density distributions of the Sn isotopes obtained from the 
RHB (me2) calculations, together with the linear functions of $r_m=0.947A^{1/3}~\textrm{fm}$, 
 $r_n=3.396+0.0191N~\textrm{fm}$, and 
$r_p=4.093+0.0069N~\textrm{fm}$, which are adjusted to the theoretical values. 
\label{fig:rmsr}}
\end{figure}

The neutron $(r_n)$ and proton $(r_p)$ rms radii of the Sn isotopes 
using the RHB (me2 and pc1) and SHFB (SKM*) calculations are shown in 
Fig.~\ref{fig:rmsr}~(a), together with the experimental data.
In all calculations, $r_p$ changes almost linearly as a function of 
the neutron number $N$ along the isotope chain, and it reproduces the experimental
data well. The theoretical values of $r_n$ also exhibit a dependence of a linear function of $N$. 
In Fig.~\ref{fig:rmsr}~(b), the neutron, proton, and matter $(r_m$) rms radii 
obtained from the me2 calculations are compared with linear functions of $N$ (or $A^{1/3}$). 
The theoretical values of $r_n$ and $r_p$ can be fitted approximately
by the linear functions $r_n=3.396+0.0191N$ and $r_p=4.093+0.0069N$, respectively, 
whereas $r_m$ is  fitted roughly by  $0.947A^{1/3}~\textrm{fm}$.
However, the experimental $r_n$ values 
obtained from the $(p,p)$ reaction at 295~MeV do not show such a smooth $N$ dependence.

\begin{figure}[!h]
\begin{center}
\includegraphics[width=0.5\textwidth]{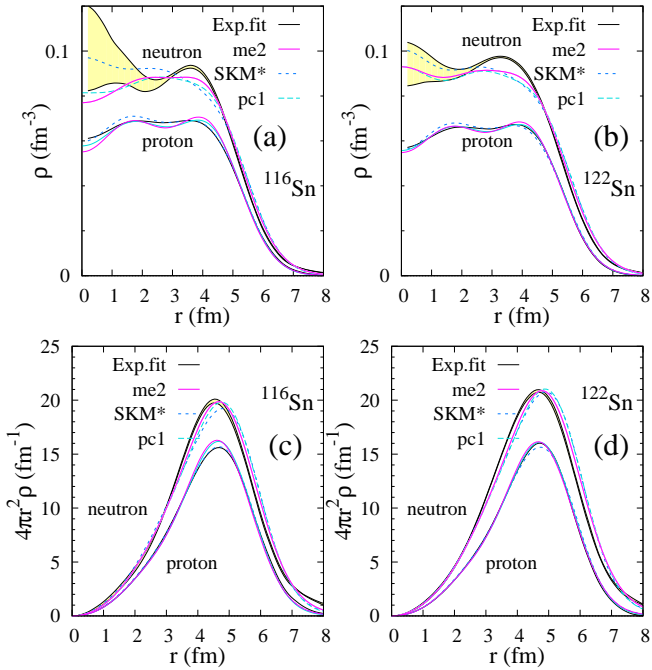}
\end{center}
  \caption{Neutron ($\rho^A_{n}$) and proton  ($\rho^A_{p}$) density distributions of $\Sn116$ and $\Sn122$. 
Panels (a) and (b) show the neutron and proton density distributions for  $\Sn116$ and $\Sn122$, respectively, 
as obtained from the me2, pc1, and SKM* calculations, and panels (c) and (d) show the corresponding values of
$4\pi r^2\rho$. 
The experimental neutron and proton densities from Ref.~\cite{Terashima:2008zza} are also shown.
The neutron density (error envelopes surrounded by thin lines) is 
extracted from proton elastic scattering at 295~MeV and the proton density (thin lines) is obtained from 
the charge distribution determined from electron elastic scattering. 
\label{fig:density}}
\end{figure}

The theoretical neutron~$(\rho_n)$ and proton~$(\rho_p)$ densities of $\Sn116$ and $\Sn122$ are shown in 
Fig.~\ref{fig:density}, together with the experimental data from Ref.~\cite{Terashima:2008zza},
in which $\rho_n$ was extracted from the $(p,p)$ reaction at 295~MeV and $\rho_p$
was obtained from the  charge density distribution determined from electron elastic scattering.
The three calculations using me2, pc1, and SKM* obtain  $\rho_p$ results approximately 
consistent with each other and that describe the experimental $\rho_p$ results 
reasonably well. However, the theoretical 
$\rho_n$ results  depend somewhat on the calculations. 
As shown in Figs.~\ref{fig:density}~(c) and \ref{fig:density}~(d), 
the position of the peak of $4\pi r^2\rho_n(r)$ is shifted outward in the calculations that used the 
pc1 and SKM* densities 
compared with those utilizing the me2 density. These differences in the surface neutron densities in the region 
$4~\textrm{fm}\lesssim r\lesssim 6~\textrm{fm}$ produce 
differences among the theoretical predictions of the $(p,p)$ cross sections at 295~MeV,
as shown later, even though the me2, pc1, and SKM* calculations give approximately consistent values of 
$r_n$.

\subsection{Sn$(p,p)$ cross sections at 295~MeV}

\begin{figure}[!h]
\includegraphics[width=0.5\textwidth]{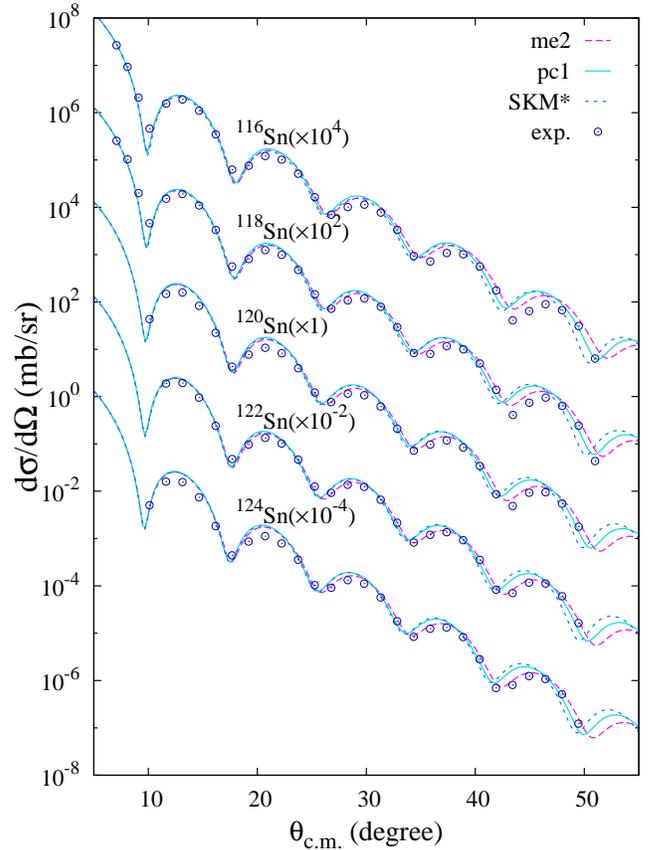}
\vskip 1.5cm
  \caption{Sn$(p,p)$ cross sections at 295~MeV obtained from the RIA+ddMH calculations using the 
theoretical densities from the me2, pc1, and SKM* calculations, together with the 
experimental data~\cite{Terashima:2008zza}. 
\label{fig:cross-skm}}
\end{figure}

Figure~\ref{fig:cross-skm} shows
the Sn$(p,p)$ cross sections at 295~MeV 
obtained from the RIA+ddMH calculations using the theoretical densities,
together with the experimental data. 
The calculated cross sections at backward angles depend on the adopted theoretical 
density.
Compared with the experimental cross sections, 
the me2 density yields reasonable agreement with the data, whereas the pc1 and SKM* densities do not;
In the results for the pc1 and SKM* densities, 
the diffraction pattern has shrunk.
because of the outward shift of the surface-peak position of $4\pi r^2\rho(r)$ explained previously;
hence, the positions of the dips at backward angles are  
shifted to forward angles and deviate significantly from the experimental data.
This indicates that the $(p,p)$ reactions at 295~MeV are good probes for determining 
the surface neutron density.

\begin{figure}[!h]
\includegraphics[width=0.5\textwidth]{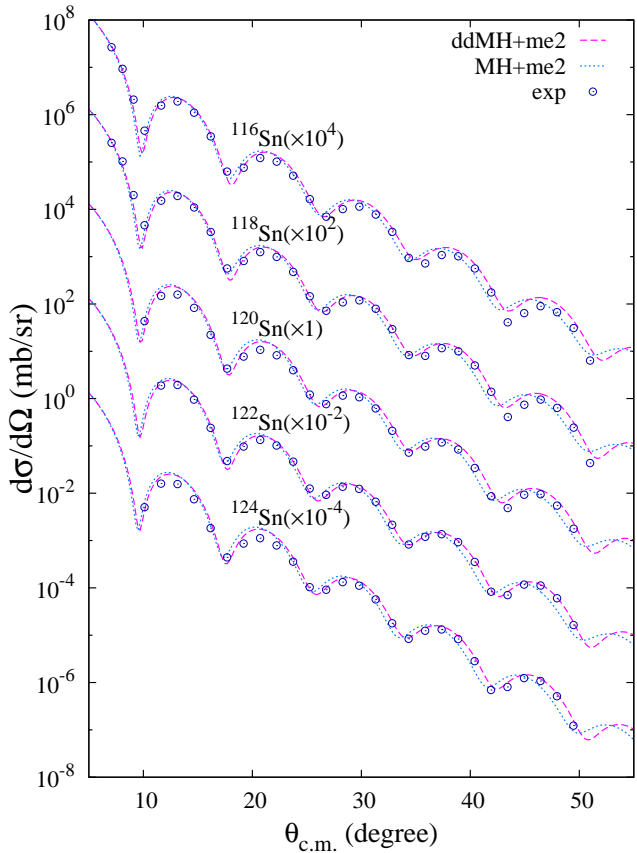}
\vskip 1.5cm 
  \caption{Sn$(p,p)$ cross sections at 295~MeV obtained from the RIA+MH calculations using the 
me2 density compared with those obtained from the RIA+ddMH calculations and 
with the experimental data~\cite{Terashima:2008zza}.
\label{fig:cross-MH}}
\end{figure}

To see the dependence of the reaction calculations on the effective $NN$ interaction model,
the present RIA+ddMH and original RIA+MH calculations with the me2 density 
are compared in Fig.~\ref{fig:cross-MH}. 
Compared with RIA+MH, the RIA+ddMH calculations obtain
a diffraction pattern that is expanded slightly toward larger angles, which means that the range of
 the  $NN$ interaction is effectively shorter in the RIA+ddMH model.

Among these theoretical calculations of the Sn$(p,p)$ reaction, 
the RIA+ddMH calculations with the me2 density obtains yield
the best agreement with the experimental cross sections. 
In particular, they reproduce 
the $\Sn122(p,p)$ cross sections fairly well. However, for the $\Sn116(p,p)$ and  $\Sn118(p,p)$ cross sections, 
the agreement at backward angles is not satisfactory. This indicates 
that a modification of the theoretical densities of $\Sn116$ and $\Sn118$ from the me2 calculation 
is necessary to reproduce the $(p,p)$ data.

\subsection{Renormalization of the experimental cross sections}
As discussed previously, 
the Sn$(p,p)$ cross sections at backward angles are sensitive to the surface neutron density 
of the target nucleus and also depend on the effective $NN$ interaction used in the reaction calculations. 
However, the model dependence at forward angles is quite small; there is almost no difference 
between the calculations in the region  
$\theta_\textrm{c.m.}\lesssim 16^\circ$,
as shown in Figs.~\ref{fig:cross-skm} and \ref{fig:cross-MH}.
All the calculations reproduce the $\Sn116(p,p)$, $\Sn118(p,p)$, and $\Sn122(p,p)$ cross sections
at forward angles reasonably well but cannot describe the $\Sn120(p,p)$ nor the $\Sn124(p,p)$ data.
In the experimental data for the 
$\Sn120(p,p)$ and $\Sn124(p,p)$ cross sections, the peak height at $\theta_\textrm{cm}\approx 13^\circ$ 
is significantly smaller than expected from the isotopic systematic than are those for the other Sn isotopes.

\begin{figure}[!h]
\includegraphics[width=8 cm]{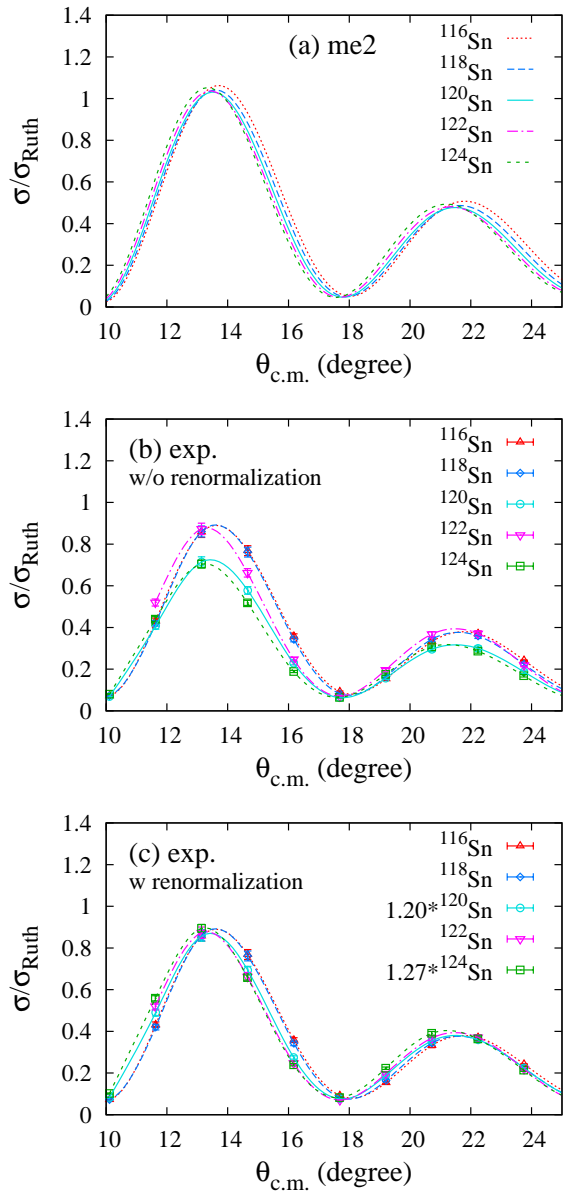}
  \caption{(a) Rutherford ratios of the Sn$(p,p)$ cross sections at 295~MeV obtained from the RIA+ddMH calculations
using the me2 density and (b) from the experimental data~\cite{Terashima:2008zza}. (c) 
The Rutherford ratio for the data from the original  
$\Sn120(p,p)$ and $\Sn124(p,p)$ cross sections of Ref.~\cite{Terashima:2008zza} 
renormalized 
by the factors 1.20 and 1.27, respectively. 
The experimental data points with error bars are connected by spline curves.
\label{fig:cross-linear}}
\end{figure}

To show this inconsistency in the data in more detail, 
the Rutherford ratios of the $(p,p)$ cross sections are compared in 
linear plots for the five Sn isotopes 
in Figs.~\ref{fig:cross-linear}~(a) and \ref{fig:cross-linear}~(b). 
The theoretical cross sections obtained from the RIA+ddMH calculation with the me2 density show 
reasonably systematic variations over the series of Sn isotopes;
the peak position changes slightly but gradually to forward angles 
from $\Sn116$ to $\Sn124$ because of the increasing nuclear size, 
but the peak height is almost unchanged. However, in the experimental data, 
the peak heights of the $\Sn120(p,p)$ and $\Sn124(p,p)$ cross sections 
are about 20\% smaller than those for the other Sn isotopes.
Because the Rutherford ratio of the cross sections at the forward-angle peak usually close
 to unity for the $(p,p)$ reactions, 
the experimental cross sections for $\Sn120(p,p)$ and $\Sn124(p,p)$ seem
unexpectedly small, indicating that the normalization of these data should be corrected.
Therefore, the $\Sn120(p,p)$ and $\Sn124(p,p)$ data were renormalized by multiplying them by 
factors of 1.20 and 1.27, respectively, to adjust the theoretical values 
$\sigma(\SnA)/\sigma(\Sn122)$ of the cross section ratios for 
$\SnA(p,p)$ to $\Sn122(p,p)$ at $\theta_\textrm{c.m.}=13.14^\circ$ obtained 
from the RIA+ddMH calculations with the me2 density.
As shown in Fig.~\ref{fig:cross-linear}~(c), 
the renormalized data for the cross sections 
show a reasonably systematic variation over the series of Sn isotopes. 
This calibration of the normalization of the $\Sn120(p,p)$ and $\Sn124(p,p)$ data using 
the theoretical prediction has almost no model 
dependence because all the calculations show quite similar isotopic systematics 
for the forward-angle cross sections; the model uncertainties in the normalization are less than a few percent. 
In the following analysis, the renormalized data are mainly used for the $\Sn120(p,p)$ and $\Sn124(p,p)$ 
cross sections.

\subsection{Isotopic properties of nuclear structure and reactions}\label{sec:isotopic}
Here, the structural properties of the Sn isotopes in the region $A=114$--$124$ 
obtained from the me2 calculations and their effects on the $(p,p)$ cross sections are discussed
 while focusing on the isotopic systematics.

\subsubsection{Structural properties}
\begin{figure}[!h]
\begin{center}
\includegraphics[width=0.5\textwidth]{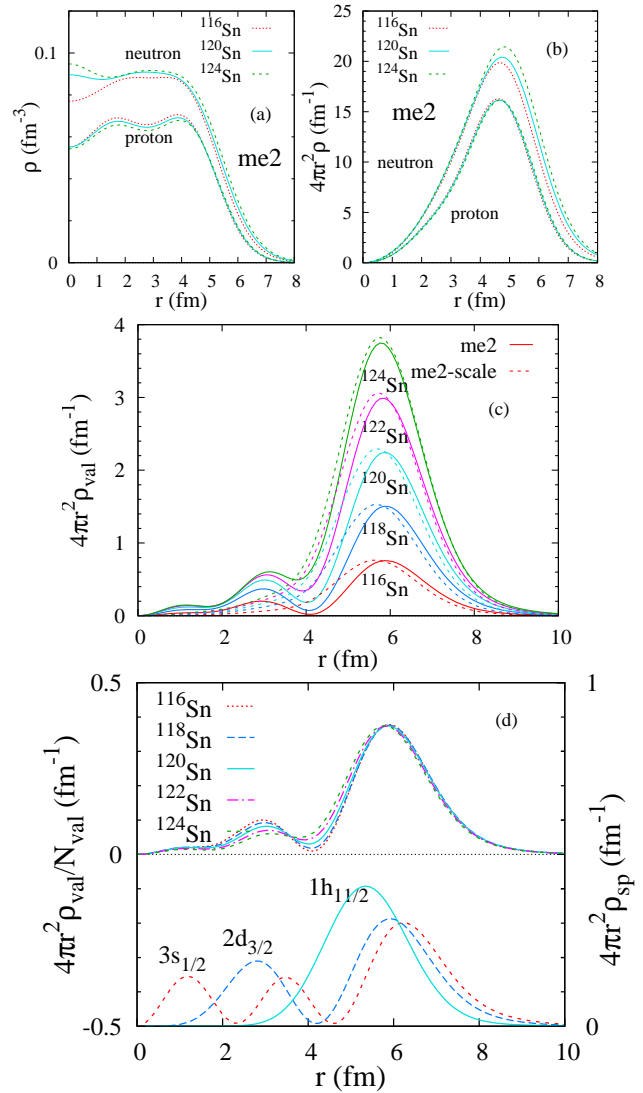}
\end{center}
  \caption{
(a) The neutron and proton density distributions
 of the Sn isotopes obtained from the me2 calculations
and  (b) the corresponding values of $4\pi r^2\rho_{n,p}(r)$.
(c) The valence neutron density $\rho_\textrm{val}\equiv \rho^A_n-\rho^{114}_n$
around the $\Sn114$ core for the me2 density.
The density $\rho^A_{n,\textrm{scale}}(r)-\rho^{114}_n(r)$
for the me2-scale density is also shown for comparison.
(d) (upper panel) The valence neutron density per neutron $\rho_\textrm{val}/N_\textrm{val}$ ($N_\textrm{val}=N-64$) 
and (lower panel) the single-particle densities $\rho^\textrm{s.p.}_\sigma(r)$ for $\sigma=3s_{1/2}$, $2d_{3/2}$, and $1h_{11/2}$
in $\Sn120$ obtained from the me2 calculations.
\label{fig:iso-density}}
\end{figure}

\begin{figure}[!h]
\includegraphics[width=7 cm]{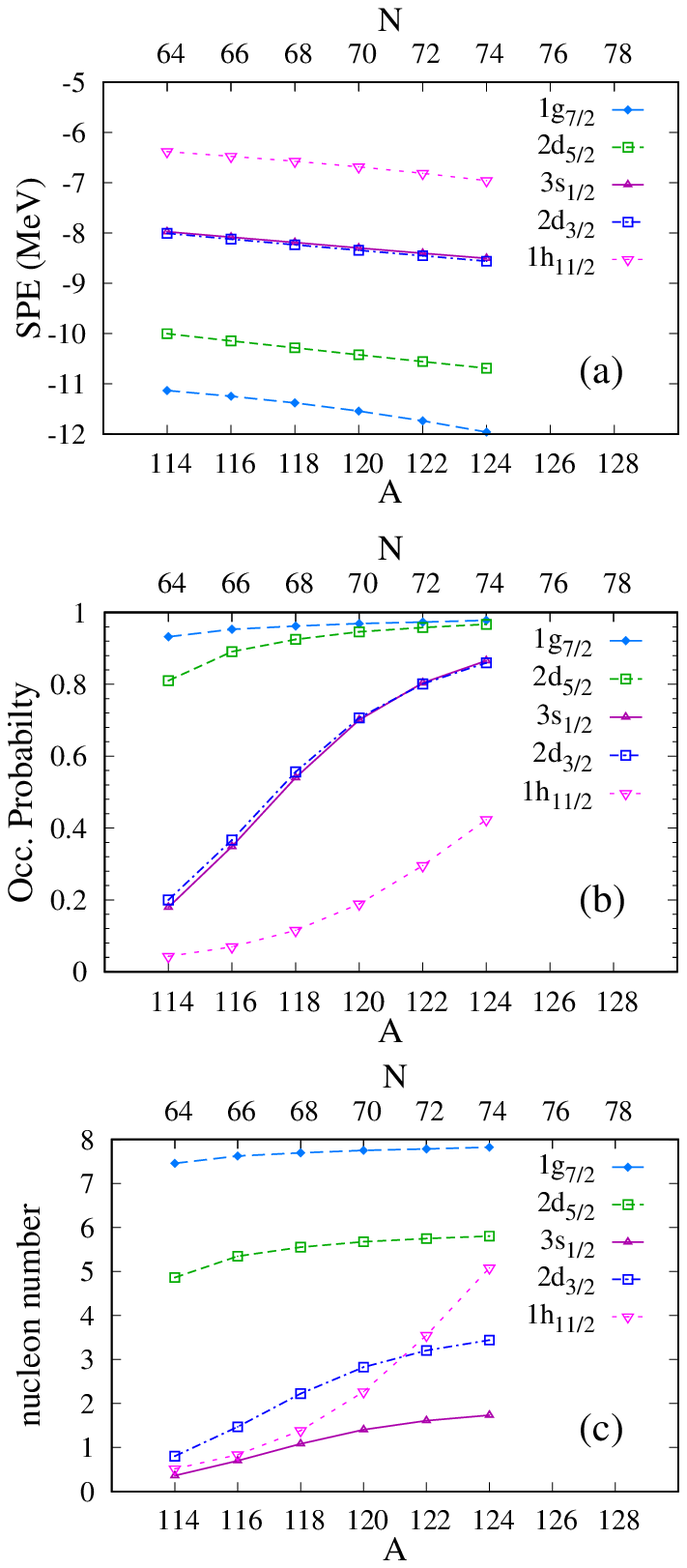}
vskip 1.5cm
  \caption{(a) Neutron single-particle energies (SPE), (b) occupation probabilities, and 
(c) the number of neutrons in the single-particle orbits in the Sn isotopes obtained from the me2 calculations. 
\label{fig:spe}}
\end{figure}

As discussed in Ref.~\cite{Bassem:2019gil}, 
RHB calculations using such interactions as me2 and pc1 show features of neutron subshell closure
at $\Sn114$ because of the neutron-shell gap at $N=64$, although the signal is not especially remarkable. 
In the range $A=116$--$124$, from $\Sn116$ to $\Sn124$, 
the valence neutrons  occupy the $3s_{1/2}$, $2d_{3/2}$, and $1h_{11/2}$ orbits in the major shell on 
top of the $\Sn114$ core. These valence-neutron orbits contribute to the 
surface neutron density in the Sn isotopes. 
The theoretical results for the proton and neutron densities and valence-neutron densities obtained from 
the me2 calculations are shown in Fig.~\ref{fig:iso-density};
Figures~\ref{fig:iso-density}~(a) and \ref{fig:iso-density}~(b) shows $\rho_{n,p}(r)$ and $4\pi r^2 \rho_{n,p}(r)$, respectively, whereas Fig.~\ref{fig:iso-density}~(c) 
shows the valence neutron density, $\rho_\textrm{val}(r)\equiv \rho^A_n(r)-\rho^{114}_n(r)$, and 
Fig.~\ref{fig:iso-density}~(d) shows the valence neutron density per neutron, $\rho_\textrm{val}/N_\textrm{val}$ with $N_\textrm{val}=N-64$.
Here $\rho^A_{n(p)}(r)$ indicates the neutron (proton) density of $\SnA$ obtained from the me2 calculation. 
The single-particle densities of the major-shell orbits---$3s_{1/2}$, $2d_{3/2}$, 
and $1h_{11/2}$---in $\Sn120$ are also shown in Fig.~\ref{fig:iso-density}~(d). The surface neutron density  changes gradually from $\Sn116$ to $\Sn124$ and 
the peak position of $4\pi\rho_n(r)$ shifts outward (in the large $r$ direction) as valence neutrons increase, 
whereas the proton density is almost unchanged [Fig.~\ref{fig:iso-density}~(a) and (b)].
The value of 
$4\pi r^2 \rho_\textrm{val}$ for the valence-neutron density 
shows an enhanced peak at $r\approx 6~\textrm{fm}$ [Fig.~\ref{fig:iso-density}~(c)].
In the valence-neutron density per neutron, $4\pi r^2 \rho_\textrm{val}/N_\textrm{val}$, 
there are small differences between isotopes in the region $r < 6~\textrm{fm}$, but there is 
almost no difference in the region $r \gtrsim 6~\textrm{fm}$ [Fig.~\ref{fig:iso-density}~(d)]. 
Such a weak $N$ dependence of $\rho_\textrm{val}/N_\textrm{val}$ is consistent with the 
valence-neutron orbits in the major shell. Figure~\ref{fig:spe} shows  
the energies, occupation probabilities, and 
neutron numbers of single-particle orbits in the Sn isotopes.
As shown in Figs.~\ref{fig:spe}~(b) and \ref{fig:spe}~(c),
the valence neutrons gradually occupy three major-shell 
orbits---$3s_{1/2}$, $2d_{3/2}$, and $1h_{11/2}$---indicating that 
shell effects are smeared by the pairing effect. 
This isotopic systematic a surface neutron density that is smoothly changing 
from $\Sn116$ to $\Sn124$ 
can be described approximately by the radial scaling of 
a reference neutron density. To demonstrate this radial-scaling property, 
scaled densities 
$\rho^A_{n,\textrm{scale}}(r)$ and $\rho^A_{p,\textrm{scale}}(r)$ 
for the neutron and proton densities of $\SnA$ were constructed 
from $\rho^{122}_{n,p}(r)$ for the reference isotope $\Sn122$;
\begin{align}
\rho^A_{n,\textrm{scale}}(r)&=\frac{N}{72}\frac{1}{s_n^3}\rho^{122}_n(r/s_n),\\
\rho^A_{p,\textrm{scale}}(r)&=\frac{1}{s_p^3}\rho^{122}_p(r/s_p),
\end{align}
where the radial scaling parameters $s_n$ and $s_p$ are chosen to be 
linear functions---$s_n=1+0.004(N-72)$ and $s_p=1+0.0015(N-72)$---that fit 
the theoretical values of $r_n$ and $r_p$, respectively, as shown in Fig.~\ref{fig:rmsr}~(b).
We label the scaled density as the me2-scale density. 
The valence-neutron density,
$4\pi r^2 \rho_\textrm{val}=\rho^A_{n,\textrm{scale}}(r)-\rho^{114}_n(r)$ ($A=116$--124),
for the me2-scale density is shown in Fig.~\ref{fig:iso-density}~(c), and 
the Sn$(p,p)$ cross sections
at 295~MeV obtained from the RIA+ddMH calculations using the me2-scale density are shown in 
Fig.~\ref{fig:cross-me-3s1}~(a), for comparison with the results using the me2 density. 
The me2-scale density describes the property of $4\pi r^2 \rho_\textrm{val}$ around the 
peak and yields $(p,p)$ cross sections that are almost equivalent to those obtained using  
the me2 density. 
This indicates that the essential features of the surface neutron density obtained from the
me2 calculations are simply described by radial scaling with a linear function.

\subsubsection{Sensitivity of $(p,p)$ cross sections to the neutron structure}

As explained previously, the RIA+ddMH calculations with the me2 density reproduce
the experimental data for the Sn$(p,p)$ cross sections at 295MeV reasonably well but 
overestimate the absolute amplitudes of the $\Sn116(p,p)$ and $\Sn118(p,p)$ cross sections at 
backward angles although they do describe the dip positions. 

\begin{figure}[!h]
\includegraphics[width=0.5\textwidth]{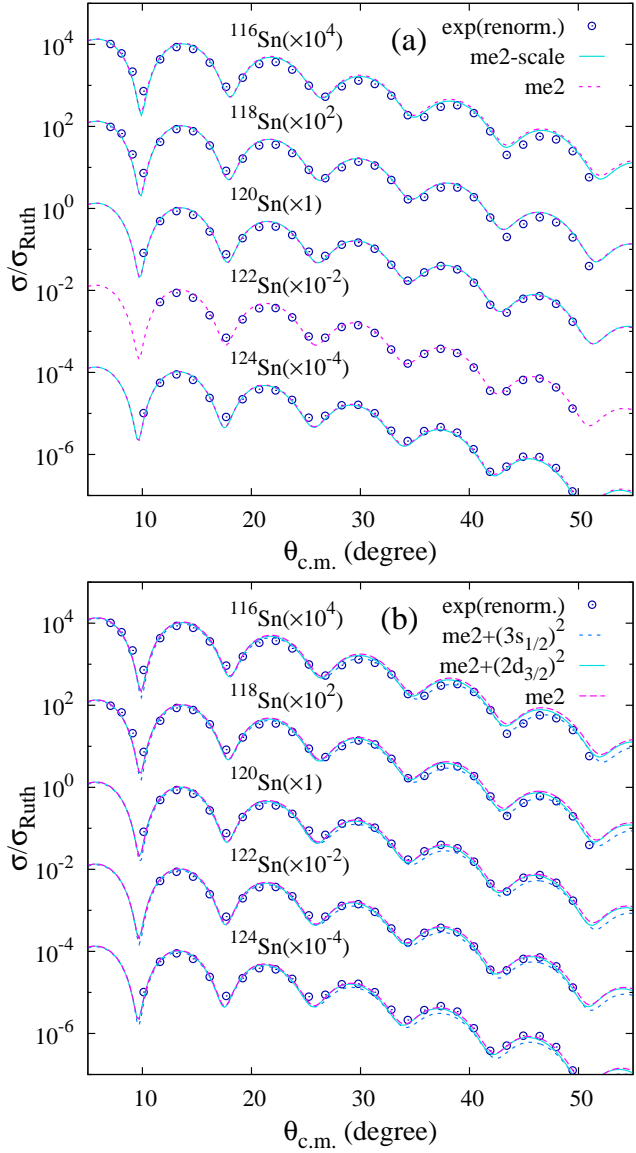}
\vskip 1.5cm
  \caption{(a) Sn$(p,p)$ cross sections at 295~MeV obtained from the RIA+ddMH calculations
using the me2 and me2-scale densities. (b) The cross sections  obtained using the modified densities for the $(3s_{1/2})^2$ and $(2d_{3/2})^2$ cases together with the results obtained using the original me2 density. 
The renormalized data for the experimental cross sections~\cite{Terashima:2008zza} are also shown; 
the $\Sn120(p,p)$ and $\Sn124(p,p)$ cross sections are renormalized from 
the original data of Ref.~\cite{Terashima:2008zza} by factors of 1.20 and 1.27, respectively.
\label{fig:cross-me-3s1}}
\end{figure}

This indicates that some
modification of the me2 density is needed for $\Sn116$ and $\Sn118$. 
The essential feature that contributes to the cross sections is 
the surface neutron density, which is sensitive to the valence-neutron orbits in the 
major shell.
To see the effects 
from each major-shell orbit to the $(p,p)$ cross sections at 295~MeV, 
modified neutron densities were constructed from the me2 density
by replacing part of the neutron density in the original me2 density for $\SnA$ 
with a two-neutron density in the $3s_{1/2}$, $2d_{3/2}$, or $1h_{11/2}$ orbit;
\begin{align}
\rho^A_n(r)&=(1-\frac{2}{N})\rho^A_n(r)+2\rho^\textrm{s.p.}_{\sigma}(r),
\end{align}
where $\rho^\textrm{s.p.}_{\sigma}(r)$, with $\sigma=\{3s_{1/2},2d_{3/2},1h_{11/2}\}$,
are the single-particle densities obtained for $\Sn120$,
which are shown in the lower panel of Fig.~\ref{fig:iso-density}~(d).
The $(p,p)$ cross sections obtained using this modified density for the $(3s_{1/2})^2$, 
 $(2d_{3/2})^2$, and  $(1h_{11/2})^2$ cases are shown in Fig.~\ref{fig:cross-me-3s1}~(b).
In the result calculated  for the $(3s_{1/2})^2$ case, the cross sections 
at backward angles are suppressed, and improved results are obtained 
for the $\Sn116(p,p)$ and  $\Sn118(p,p)$ cross sections compared with the original me2 results. 
However, in the results for the 
$(2d_{3/2})^2$ and  $(1h_{11/2})^2$ cases, the cross sections are almost unchanged from the 
me2 result.
This means that the cross sections are sensitive to $3s_{1/2}$ neutrons,
as expected from the general trend of higher nodal orbits that provide significant 
contributions to the surface neutron density.


\clearpage

\section{Isotopic analysis of cross sections using model density}\label{sec:analysis}


To fine tune the neutron density to fit the $(p,p)$ cross section data, 
I introduce a model density that extends the modified density with $(3s_{1/2})^2$ neutrons.
This corresponds to a modification of the me2 density and is called the me2-3s model.
In the me2-3s model, the neutron density $\rho^A_{n,3s}(r)$ is expressed as
\begin{align}
&\rho^A_{n,3s}(r)=\frac{1}{(1+\delta_s)^3}\cdot\nonumber\\
&\Bigl\{\left(1-\frac{v}{N}\right)
\rho^A_n\left(\frac{r}{1+\delta_s}\right)+v\rho^\textrm{s.p.}_{3s_{1/2}}\left(\frac{r}{1+\delta_s}\right)
\Bigr\}
\end{align}
with the two parameters $v$ and $\delta_s$ that respectively indicate the enhancement of the $3s_{1/2}$ neutron density
and the radial~$(r)$ scaling, respectively. 
The case $v=2$ cwith $\delta_s=0$ (no scaling) corresponds
to the modified density for the $(3s_{1/2})^2$ case from the previous discussion. 
Note that the parameter $v$ does not directly correspond to an increase in the density of $3s_{1/2}$ neutrons, 
but it effectively controls the contribution of the $3s_{1/2}$ neutrons to the neutron density.  
The isotopic analysis of $\SnA(p,p)$ at 295~MeV was performed using
the RIA+ddMH  calculation with the 
me2-3s density.


For the isotopic analyses of the neutron density and $(p,p)$ cross sections, 
the isotopic neutron density difference and the isotopic cross section ratio are
defined by adopting $\Sn122$ as the reference isotope;
\begin{align}
&D(\rho_{n};r)\equiv \rho_{n}(\SnA;r)-\rho_{n}(\Sn122;r),\\
& R(\sigma;\theta_\textrm{c.m.})\equiv \frac{d\sigma(\SnA)/d\Omega}{d\sigma(\Sn122)/d\Omega},
\end{align}
where $d\sigma(\SnA)/d\Omega$ are the differential cross sections for the $\SnA(p,p)$ reactions 
in the center-of-mass frame.
The experimental values of $R(\sigma;\theta_\textrm{c.m.})$ are obtained
from the Sn$(p,p)$ cross section data measured at the same angles in the laboratory frame
omitting a slight difference in the angles in the center-of-mass frame caused by the mass difference between
isotopes.

\begin{figure}[!h]
\includegraphics[width=0.5\textwidth]{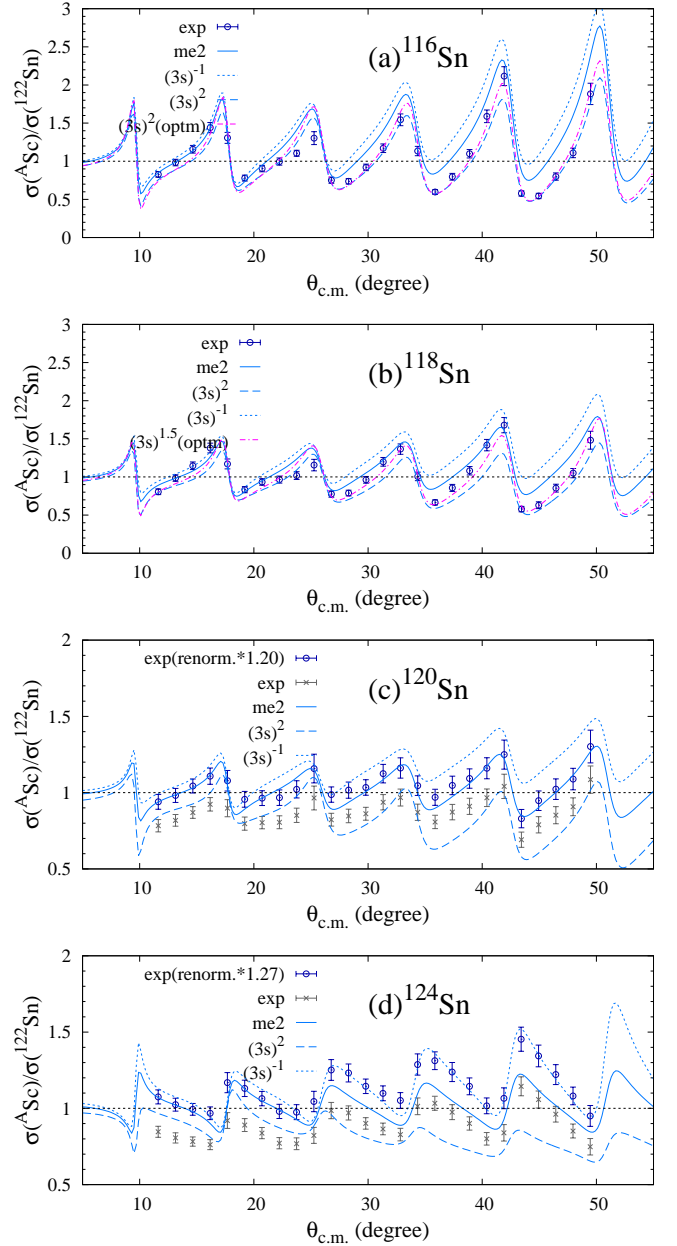}
  \caption{Isotopic cross section ratio $R(\sigma)$ for $\SnA(p,p)$ to $\Sn122(p,p)$ at 295~MeV 
obtained from the RIA+ddMH calculations using the me2-3s density with 
$(v,\delta_s)=(2,0)$ and $(-1,0)$ [labeled $(3s)^2$ and $(3s)^{-1}$, respectively]. 
In panels (a) and (b)
for $\Sn116(p,p)$ and $\Sn118(p,p)$, the results obtained using the me2-3s density
with the optimized parameter sets $(v,\delta_s)=(2,-0.7\%)$ and $(v,\delta_s)=(1.5,-0.7\%)$ are 
also shown, with the labels 
$(3s)^{2}$(optm) and $(3s)^{1.5}$(optm), respectively.   
The experimental values of $R(\sigma)$ include those obtained using the original data of Ref.~\cite{Terashima:2008zza}
and those obtained using the data for the $\Sn120(p,p)$ and $\Sn120(p,p)$ cross sections renormalized 
by the factors 1.20 and 1.27, respectively. 
\label{fig:cross-compare}}
\end{figure}

Figure~\ref{fig:cross-compare} shows 
the isotopic cross section ratio $R(\sigma)$ for $\SnA(p,p)$ at 295~MeV 
calculated using the me2 and me2-3s densities,
together with the experimental values obtained from the
$(p,p)$ cross section data. 
Here, the result obtained using the me2 density, which is shown by the solid lines,
is first discussed. 
The ratio $R(\sigma)$ shows an oscillating behavior that corresponds to a slight 
shift in the diffraction pattern of the cross sections from $\Sn122$ to $\SnA$, 
which probes the change in nuclear size via the nucleon-nucleus optical potentials. 
As discussed in the previous analysis of Pb isotopes~\cite{Kanada-Enyo:2021}, the oscillation amplitude of $R(\sigma)$ 
is dominantly determined by the isotopic neutron-radius difference, whereas the gradual deviation 
from the line $R=1$ is sensitive to the detailed profiles of the surface neutron density around the peak of $4\pi r^2\rho_n(r)$. 
As $N$ decreases from $\Sn122$ to $\Sn116$, the oscillation amplitude of $R(\sigma)$ increases because of the shrinking nuclear size. The opposite oscillation pattern of $R(\sigma)$ obtained for $\Sn124$ indicates 
that the nuclear size increases from $\Sn122$ to $\Sn124$, but the oscillation amplitude is 
consistent with that for $\Sn120$ because the size difference 
from the reference isotope $\Sn122$ has almost the same magnitude for $\Sn120$ and $\Sn124$.

Next, let me turn to the results obtained using the me2-3s model.  In Fig.~\ref{fig:cross-compare}, 
the ratio $R(\sigma)$ obtained using the me2-3s density with $v=2$ and $v=-1$
for the case $\delta_s=0$ (no scaling) are shown
by dashed and dotted lines, which respectively correspond to enhanced and suppressed $3s_{1/2}$ neutron components. 
In the result for $\Sn116$, the me2-3s model with $(v,\delta_s)=(2,0)$ yields
a better agreement than the original me2 result, 
but a slight disagreement with the experimental data still remains.
By tuning the radial scaling parameter $\delta_s$, 
the me2-3s density with the parameter set 
$v=2$ and $\delta_s=-0.7\%$ was obtained as an optimized solution to reproduce the 
experimental $R(\sigma)$ for $\Sn116$. 
A similar analysis was performed for $\Sn118$,
and the optimized parameter set $(v,\delta_s=(1.5,-0.7\%)$ was 
obtained to describe the experimental $R(\sigma)$.

For $\Sn120$, 
the $R(\sigma)$ values obtained from
the renormalized data for the $\Sn120(p,p)$ cross sections are successfully described by the 
calculations using the me2 density,
indicating that no correction to the original me2 density is necessary.
For $\Sn124$, the $R(\sigma)$ values 
obtained from the renormalized data for the $\Sn124(p,p)$ cross sections can be 
described by the me2-3s density with 
$(v,\delta_s)=(-1,0)$, which corresponds to a slight decrease in the $3s_{1/2}$ neutron density
from the original me2 density. 
Note that the experimental $R(\sigma)$ obtained from the original $\Sn120(p,p)$ and $\Sn124(p,p)$ 
data without renormalization deviates significantly from $R=1$, and it is difficult to be described it with 
these calculations. 

\begin{figure}[!h]
\includegraphics[width=0.5\textwidth]{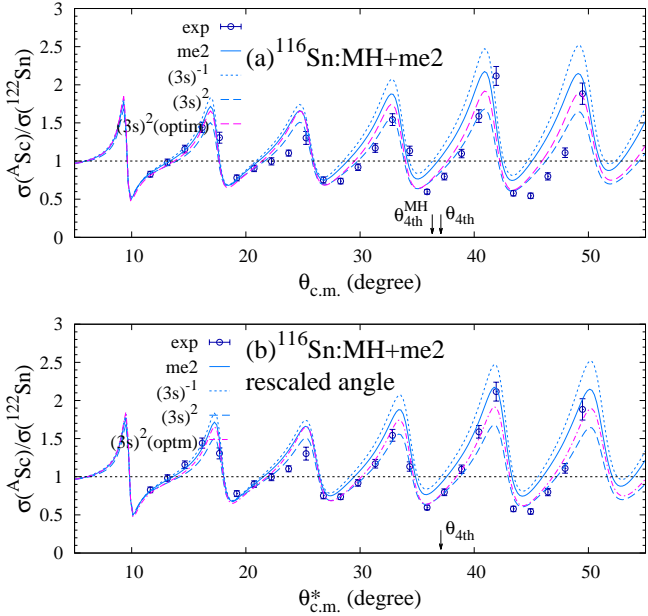}
  \caption{
(a) Same as Fig.~\ref{fig:cross-compare}(a) but using the results
obtained from the RIA+MH calculations together with the experimental values.
(b) The same ratios plotted against the rescaled angles $\theta_\textrm{c.m.}^*=S_\theta \theta_\textrm{c.m.}$ obtained with
the angle-scaling factor $S_\theta=\theta_\textrm{4th}/\theta^\textrm{MH}_\textrm{4th}$.
The angle $\theta_\textrm{4th}$ ($\theta^\textrm{MH}_\textrm{4th}$)
of the fourth peak of the  $\Sn122(p,p)$ cross sections obtained from the RIA+ddMH (RIA+MH) calculations
using the me2 density is shown by the arrows.
The experimental values obtained from the cross section data of Ref.~\cite{Terashima:2008zza} are plotted for 
$\theta_\textrm{c.m.}$.
\label{fig:cross-compare-MH}}
\end{figure}

To discuss the dependence of the present analysis of $R(\sigma)$ on the effective $NN$ 
interaction model used in the reaction calculations, I perform the RIA+MH calculations using the me2 and me2-3s densities 
and compare the result with the RIA+ddMH calculations. As shown in Fig.~\ref{fig:cross-compare-MH}~(a),
the oscillation interval of $R(\sigma)$ is slightly shorter for the RIA+MH result than for RIA+ddMH.
However, by rescaling the angles $\theta\to \theta^*=S_\theta\theta_\textrm{c.m.}$, 
almost consistent results are obtained for both calculations. Here 
the angle-scaling factor is chosen to be $S_\theta=\theta_\textrm{4th}/\theta^\textrm{MH}_\textrm{4th}$
so as to fit the  angle ($\theta^\textrm{MH}_\textrm{4th}$) of the fourth peak of the 
 $\Sn122(p,p)$ cross sections obtained  from  the MH calculations to that of the ddMH calculations ($\theta_\textrm{4th}$). 
The $\theta^*$ plot of $R(\sigma)$ obtained from the RIA+MH calculations is 
shown in Fig.~\ref{fig:cross-compare-MH}~(b).
The optimized me2-3s density with 
$v=2$ and $\delta=0.7\%$ yields good agreement with the experimental values of $R(\sigma)$ for 
$\Sn116$, indicating that the optimized neutron density can be extracted 
with less model uncertainty by fitting the experimental values of $R(\sigma)$ in the $\theta^*$ plot.

\subsection{Structure and reaction properties with the optimized me2-3s model density}

I call the set of neutron densities obtained using the me2-3s model with the optimized parameters $(v,\delta_s)=(2,-0.7\%)$, $(1,5,-0.7\%)$, 
$(0,0)$, $(0,0)$, and $(-1,0)$ for $\Sn116$, $\Sn118$, $\Sn120$, $\Sn122$, and $\Sn124$, respectively, 
the ``me2-3s(optm)'' densities, which are obtained in the present analysis by fitting the experimental $R(\sigma)$.
Note that, for $\Sn120$ and $\Sn122$, the me2-3s(optm) density is consistent with the original me2 density
that reproduces the cross section data without modification. In this section, I discuss 
the Sn$(p,p)$ cross sections and the neutron structure properties obtained 
using the me2-3s(optm) density. 

\subsubsection{Cross sections and analyzing powers}

\begin{figure}[!h]
\includegraphics[width=0.5\textwidth]{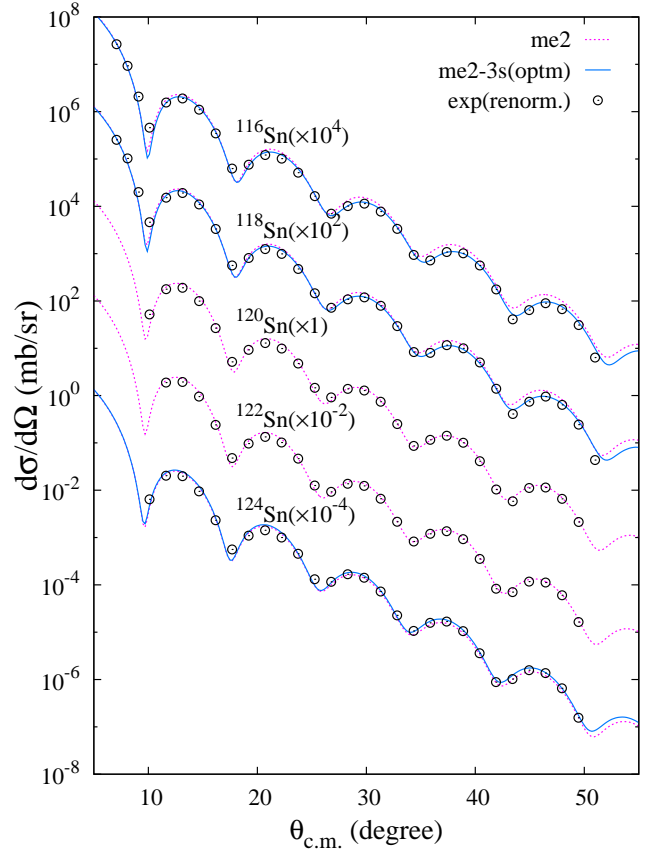}
\vskip 1.5cm
  \caption{Cross sections of Sn$(p,p)$ reactions at 295~MeV obtained from the RIA+ddMH calculations using 
the me2 and me2-3s(optm) densities together with the experimental data. 
The experimental data for the $\Sn120(p,p)$ and $\Sn124(p,p)$ cross sections are renormalized from 
the original data of Ref.~\cite{Terashima:2008zza} by factors of 1.20 and 1.27, respectively.
\label{fig:cross-mod}}
\end{figure}

\begin{figure}[!h]
\includegraphics[width=0.5\textwidth]{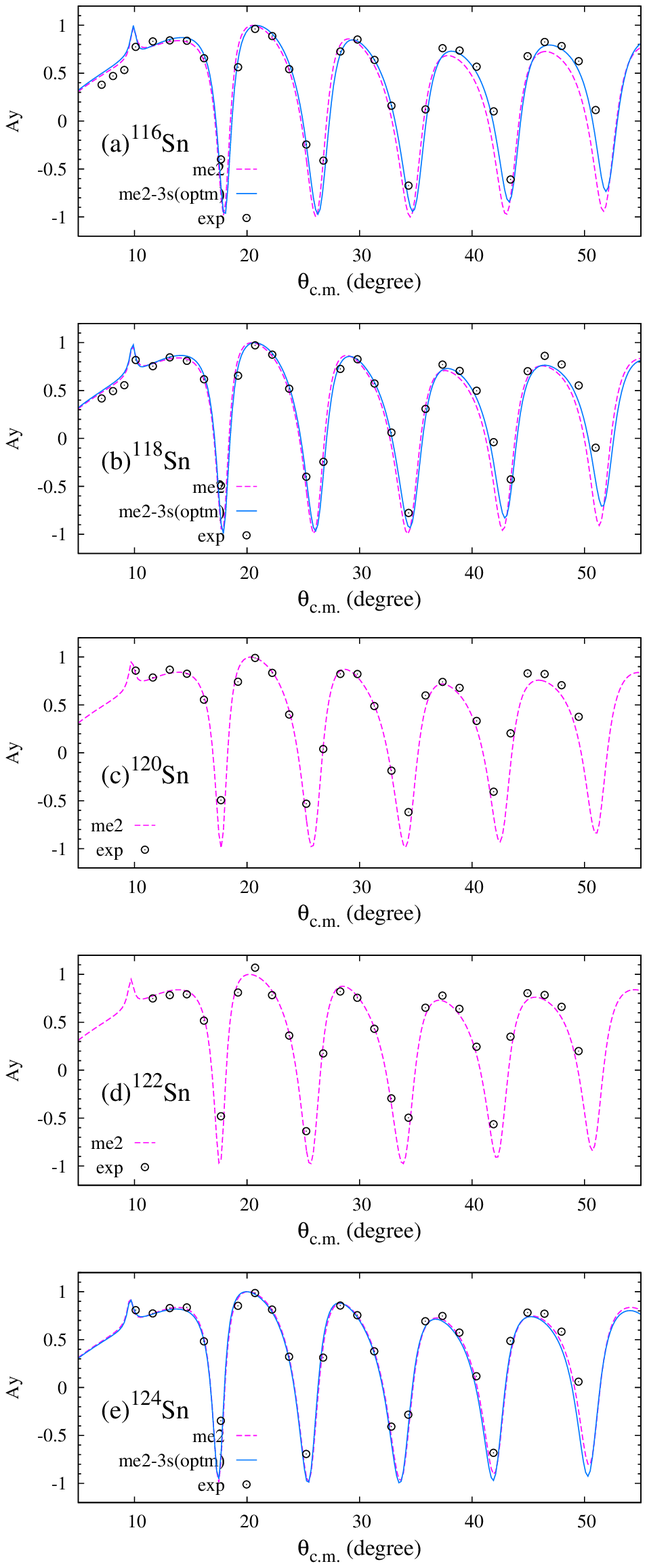}
  \caption{Analyzing powers of Sn$(p,p)$ at 295~MeV obtained from the RIA+ddMH calculations using 
the me2 and me2-3s(optm) densities together with the 
experimental data \cite{Terashima:2008zza}.
\label{fig:Ay-mod}}
\end{figure}

The cross sections and analyzing powers of Sn$(p,p)$ at 295~MeV calculated 
with RIA+ddMH using the me2-3s(opt) density are shown in Figs.~\ref{fig:cross-mod} and \ref{fig:Ay-mod},
respectively, in comparison with the experimental data and with the theoretical result obtained using the me2 density.
The present me2-3s(optm) density successfully describes the experimental $(p,p)$ data 
for the series of Sn isotopes from $\Sn116$ to $\Sn124$. 
In particular, the reproduction of the $\Sn116(p,p)$ and $\Sn118(p,p)$ cross sections at backward angles is
substantially improved by the modification from the me2 density to the me2-3s(optm) density.

\subsubsection{Neutron rms radii and skin thickness}

\begin{figure}[!h]
\includegraphics[width=8 cm]{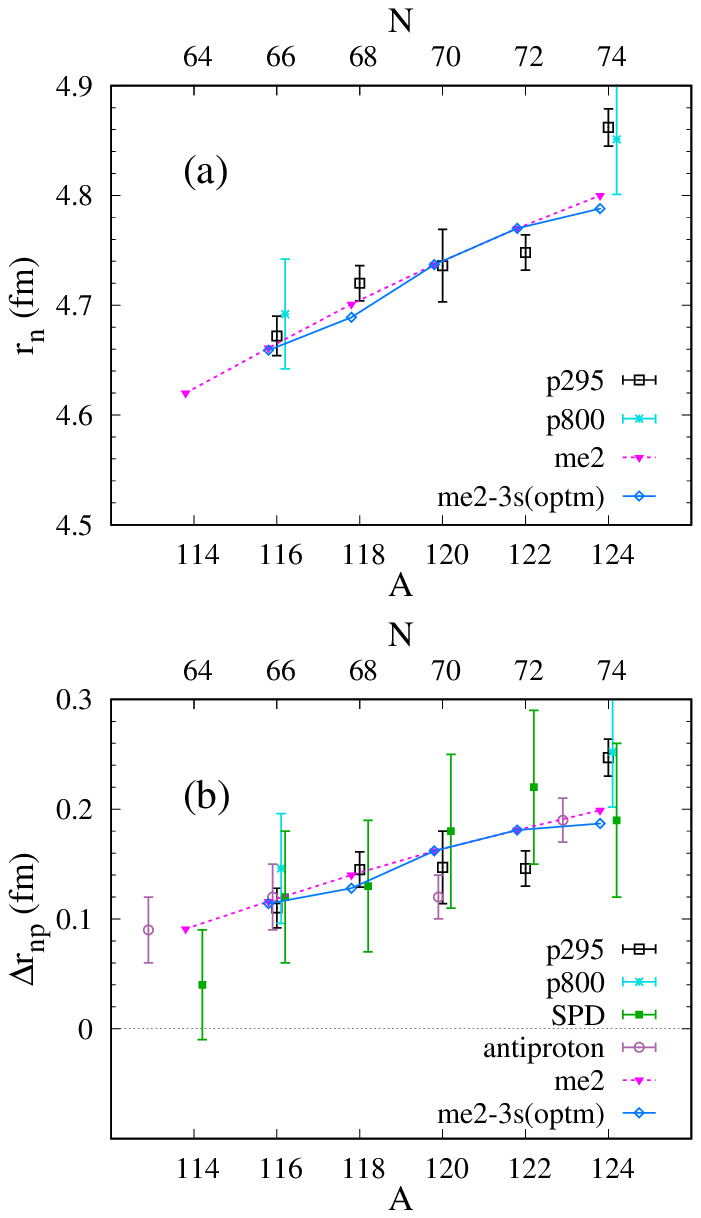}
  \caption{(a) Neutron rms radii and (b) skin thicknesses of  the Sn isotopes 
obtained from the me2 calculatiosn and those obtained using the me2-3s(optm) density, 
together with the experimental data.
The experimental data include the values of $r_n$ and $\Delta r_{np}$ 
obtained from the $(p,p)$ reactions at both 295~MeV~\cite{Terashima:2008zza} and 800~MeV~\cite{Ray:1979qv},
as well as the $\Delta r_{np}$ values obtained from 
 $x$-ray data from antiprotonic atoms~\cite{Trzcinska:2001sy} and from
spin-dipole resonances~(SPD) measured using the $(^3\textrm{He},t)$ charge-exchange reaction 
at 450 MeV~\cite{Krasznahorkay:1999zz}.
\label{fig:rmsr-mod}}
\end{figure}

The neutron rms radii $r_n$ and skin thicknesses $\Delta r_{np}$ of the Sn isotopes 
obtained for the me2-3s(optm) density are shown in 
Fig.~\ref{fig:rmsr-mod}, together with other theoretical predictions and experimental values. 
The me2-3s(optm) density gives  $r_n$ and $\Delta r_{np}$ values that are almost consistent with those of the 
me2 calculation, meaning that the modification of the surface neutron density from the me2 
to me2-3s(optm) densities  
does not make an essential contribution to the neutron rms radii, although it affects the $(p,p)$ reactions.
The obtained values show smooth $N$ dependences for $r_n$ and 
$\Delta r_{np}$. This smooth increase in $\Delta r_{np}$ with $N$ increasing is consistent with an 
theoretical work for the microscopic description of the Sn$(p,p)$ reactions at 295~MeV\cite{Haider:2010zz}, 
but seems to somewhat contradict the experimental $N$ dependences of Ref.~\cite{Terashima:2008zza}. 

\subsubsection{Neutron density and shell structure}

\begin{figure}[!h]
\begin{center}
\includegraphics[width=0.5\textwidth]{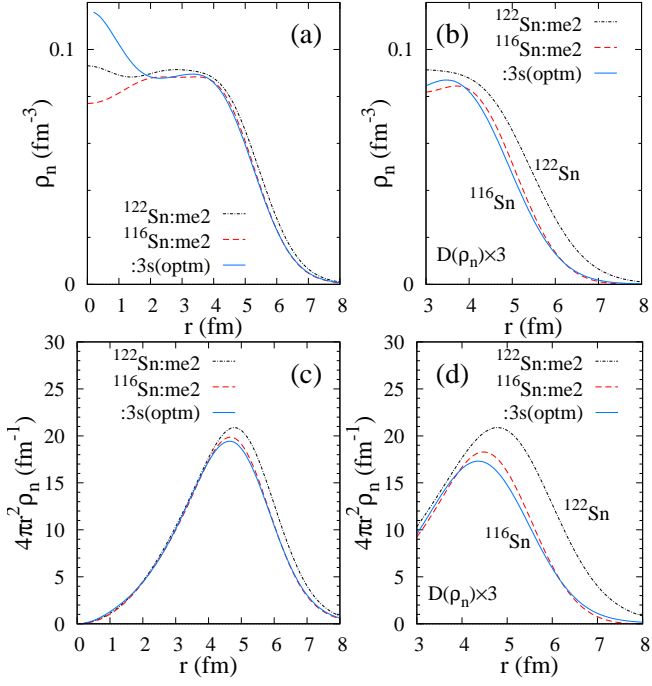}
\end{center}
  \caption{(a)~Neutron density distributions of $^{116}$Sn obtained using the me2 and me2-3s(optm)  
model densities compared with the $^{122}$Sn density. 
(b)~The density $\rho^{122}_n(r)-3D(\rho_n)$ corresponds to three times enhancement of the isotopic neutron difference $D(\rho_n)$ of $\Sn116$ 
from $\Sn122$. 
Panels (c) and (d) show the corresponding values of  $4\pi r^2\rho_n(r)$ for (a) and (b), respectively. 
but $4\pi r^2\rho_n(r)$.
\label{fig:dens-mod-116}}
\end{figure}


\begin{figure}[!h]
\begin{center}
\includegraphics[width=7cm]{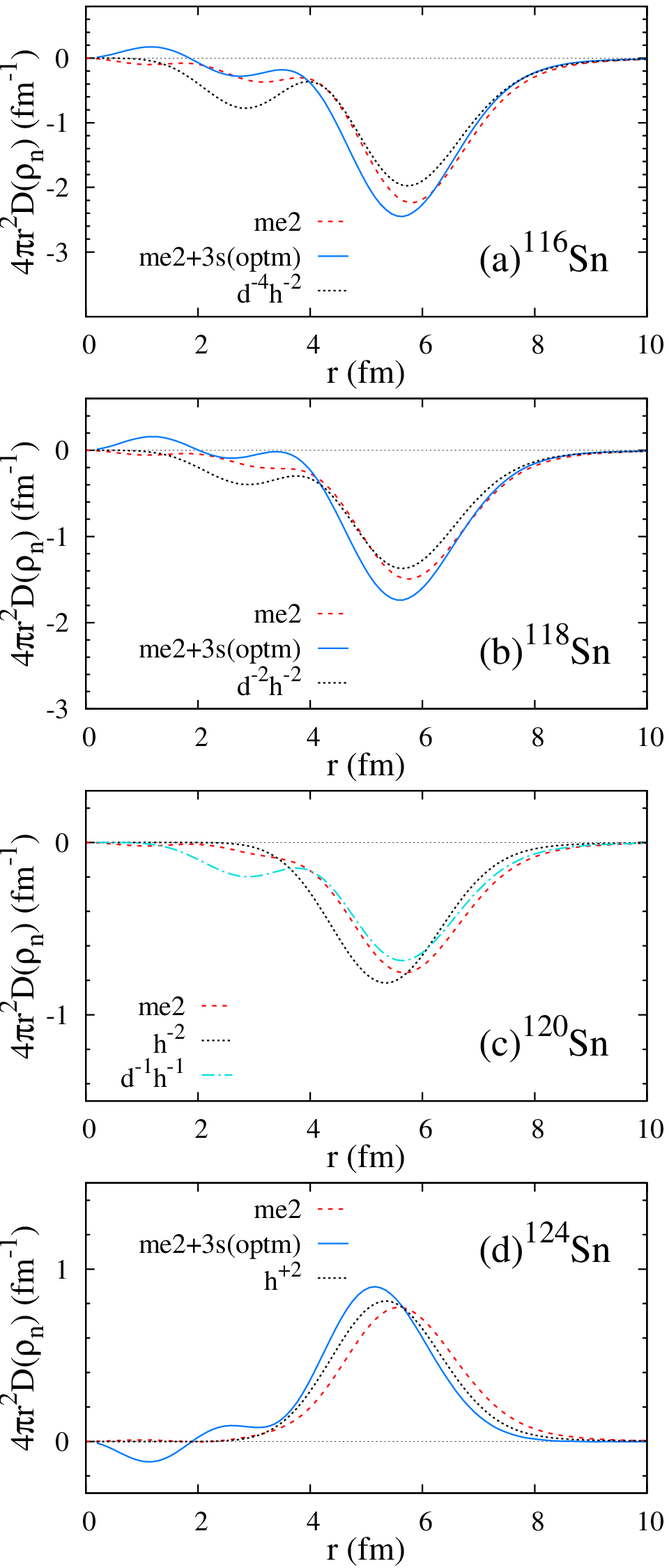}
\end{center}
  \caption{Isotopic neutron-density difference of $\SnA$ from $\Sn122$ obtained for the me2 and me2-3s(optm) densities.
Those for the lowest configurations are also shown. The configurations are listed in Table~\ref{tab:config}. 
Panel (c) also shows the hole density for the $(2d_{3/2})^{-1}(1h_{11/2})^{-1}$ configuration.
\label{fig:dens-iso-mod}}
\end{figure}
Next, the detailed properties of the neutron densities of the Sn isotopes obtained 
using the me2-3s(optm) density are discussed.
Figure~\ref{fig:dens-mod-116} compares the values of 
$\rho^{116}_n(r)$ for the me2-3s(opt) density  with 
$\rho^{116}_n(r)$ and $\rho^{122}_n(r)$ for the me2 density.
Figures~\ref{fig:dens-mod-116}~(a) 
and \ref{fig:dens-mod-116}~(c), respectively, plot the quantities 
$\rho_n(r)$ and $4\pi r^2 \rho_n(r)$.
To demonstrate the difference between the me2 and me2-3s(opt) densities more clearly, 
the densities $\rho^{122}_n(r)-3D(\rho_n)$ and $4\pi r^2(\rho^{122}_n(r)-3D(\rho_n))$
are shown in Figs.~\ref{fig:dens-mod-116}~(b) and \ref{fig:dens-mod-116}~(d), respectively, 
which correspond to three times enhancement of the isotopic neutron-density 
difference $D(\rho_n)$ of $\Sn116$ 
from $\Sn122$. As shown in Figs.~\ref{fig:dens-mod-116}~(c) and \ref{fig:dens-mod-116}~(d), 
the peak position of $4\pi r^2 \rho^{116}_n(r)$ obtained using the me2 density is shifted inward
compared with that
of $4\pi r^2 \rho^{122}_n(r)$. This change in the surface neutron density 
from $\Sn116$ to $\Sn122$ is described by the radial scaling $r\to r/s_n$ 
discussed previously. In the plot of $4\pi r^2 \rho^{116}_n(r)$ for the me2-3s(optm) density, 
the surface-peak amplitude in the region $4~\textrm{fm}\lesssim r \lesssim 6~\textrm{fm}$
is reduced from that obtained with the me2 density, whereas the peak position is almost unchanged.
This reduction of the neutron density around the surface peak
decreases the backward cross sections and improves the agreement with the
experimental $\Sn116(p,p)$ data. 

In Fig.~\ref{fig:dens-iso-mod}, 
the isotopic neutron-density differences $D(\rho_n)$ of $\SnA$ from $\Sn122$ 
in the me2 and me2-3s(optm) densities are shown by dashed and solid lines, respectively.
The correction of the surface neutron density 
from me2 to me2-3s(optm) can be clearly seen in the comparisons of $D(\rho_n)$.  
As shown in Figs.~\ref{fig:dens-iso-mod}~(a) and \ref{fig:dens-iso-mod}~(b) for $\Sn116$ and $\Sn118$,  
the amplitude of $4\pi r^2 D(\rho_n)$ in the region $4~\textrm{fm}\lesssim r \lesssim 6~\textrm{fm}$
 increases, 
and the shape of the peak changes, in going from the original me2 to the me2-3s(optm) densities.
This correction is essential for fitting the backward cross sections for the $\Sn116(p,p)$ reaction because 
$R(\sigma)$ is sensitive to $4\pi r^2 D(\rho_n)$ around the peak.
The modification of the surface neutron density in the region $4~\textrm{fm}\lesssim r \lesssim 6~\textrm{fm}$ is described
by the increase in the $3s_{1/2}$ neutron component,
as is easily understood from the nodal structure of the $3s_{1/2}$ orbit.
Indeed, in the simple case of the me2-3s model with $\delta_s=0$, 
the change of $\rho_n(r)$ from the me2 density to the me2-3s density can be written as
\begin{align}
&\rho_{n,3s}(r)-\rho_n(r)=D(\rho_{n,3s})-D(\rho_n)\nonumber\\
&=v( \rho^\textrm{s.p.}_{3s_{1/2}}(r) - \rho^A_{n}(r)/N).
\end{align}
By thus combining the isotopic analyses of nuclear structure and reaction, 
it is concluded that the suppression of the $\Sn116(p,p)$ and $\Sn118(p,p)$ cross sections at backward angles 
is the signal of an enhanced $3s_{1/2}$ neutron component, which is probed via its contribution to 
the surface neutron density and indicates the shell effect at $N=66$ in the Sn isotopes.

\begin{table}[!ht]
\caption{Lowest configurations in the major shell, with the level ordering 
\{$3s_{1/2}$, $2d_{3/2}$, $1h_{11/2}$\}, for $\Sn116$,  $\Sn118$, 
$\Sn120$,  $\Sn122$, and  $\Sn124$ and the corresponding hole configurations 
for the reference isotope $\Sn122$.
 \label{tab:config}
}
\begin{center}
\begin{tabular}{lccccccccc}
\hline
\hline
Isotopes & Lowest config.& Hole config. \\
$\Sn116$ & $(3s_{1/2})^2$ & $(2d_{3/2})^{-4}(1h_{11/2})^{-2}$\\
$\Sn118$ & $(3s_{1/2})^2(2d_{3/2})^2$ & $(2d_{3/2})^{-2}(1h_{11/2})^{-2}$ \\
$\Sn120$ & $(3s_{1/2})^2(2d_{3/2})^4$ & $(1h_{11/2})^{-2}$ \\
$\Sn122$ & $(3s_{1/2})^2(2d_{3/2})^4(1h_{11/2})^2$ & reference \\
$\Sn124$ & $(3s_{1/2})^2(2d_{3/2})^4(1h_{11/2})^4$ & $(1h_{11/2})^{2}$ \\
\hline
\hline
\end{tabular}
\end{center}
\end{table}

To discuss the shell effect, I consider 
the lowest configurations in the major shell, with the level ordering 
\{$3s_{1/2}$, $2d_{3/2}$, $1h_{11/2}$\}, without pairing. 
The lowest configurations for $\Sn116$,  $\Sn118$, 
$\Sn120$,  $\Sn122$, and  $\Sn124$ are listed in Table~\ref{tab:config},
together with the corresponding hole configurations for the reference $\Sn122$ state.
In Fig.~\ref{fig:dens-iso-mod}, the values of 
$4\pi r^2 D(\rho_n)$ for the lowest configurations are shown by dotted lines, which 
are perturbatively calculated using $\rho^\textrm{s.p.}_\sigma(r)$ in $\Sn120$.
As shown in Figs.~\ref{fig:dens-iso-mod}~(a) and \ref{fig:dens-iso-mod}~(b), 
the lowest configurations of 
$(3s_{1/2})^2$ and $(3s_{1/2})^2(2d_{3/2})^2$ for $\Sn116$ and $\Sn118$, respectively, 
can describe the peak shape of $4\pi r^2 D(\rho_n)$ obtained for the me2-3s(optm) density,
although they slightly underestimate the overall factor by about 20\%.
This result again supports the shell effect at $N=66$ in the Sn isotopes. 
Strictly speaking, however, the present result cannot exclude the possibility of a vanishing shell gap at $N=64$ between the $3s_{1/2}$ 
and $(1g_{7/2},2d_{5/2})$ orbits. To confirm this, experimental data that include the $\Sn114(p,p)$ reaction 
are needed.

For $\Sn120$, the lowest configuration fails to describe the structure of 
$4\pi r^2 D(\rho_n)$ with the me2 density. Alternatively, 
the hole configuration  $(2d_{3/2})^{-1} (1h_{3/2})^{-1}$ can describe 
the feature of $4\pi r^2 D(\rho_n)$, which 
means that configuration mixing between $2d_{3/2}$ and $1h_{11/2}$ may exist at $N\sim 70$. 
In the present me2-3s model, discussing details of the 
occupation probability for each orbit is difficult because 
higher-order effects from other orbits are effectively renormalized in the two parameters $v$ and $\delta_s$. 
Moreover, the $(p,p)$ reaction at 295~MeV is insensitive to the $1d_{3/2}$ and $1h_{11/2}$ 
neutron components as discussed previously.

\section{Summary}\label{sec:summary}
Proton elastic scattering at 295~MeV off Sn isotopes in the range $A=116$--124
was investigated using the RIA+ddMH model 
with theoretical densities for the Sn isotopes obtained from both RHB and SHFB
calculations of spherical nuclei.
The isotopic systematics of the nuclear structure 
and the reaction cross sections were investigated for the series of Sn isotopes.
The theoretical results from the structure calculations show 
a smooth $A$ dependence of the neutron rms radii and surface densities, 
along the isotope chain because of the pairing effect.
The RIA+ddMH calculations 
using the theoretical density from the RHB calculations with the me2 interaction (the so called DD-ME2
interaction) reasonably agreed 
with the experimental cross sections and analyzing powers---in particular, 
for the $\Sn122(p,p)$ reaction---
but they overestimated the backward cross sections 
of the $\Sn116(p,p)$ and $\Sn118(p,p)$ reactions.

To obtain the optimized neutron density from the experimental Sn$(p,p)$ data at 295~MeV, 
an isotopic analysis was performed based on the RIA+ddMH calculations using the me2-3s model,  in which 
the original me2 density was modified by
changing the $3s_{1/2}$ neutron component. 
The increase in the $3s_{1/2}$ neutron component made a significant contribution to the surface neutron density
and improved the agreement with the experimental data for the  $\Sn116(p,p)$ and $\Sn118(p,p)$ cross sections  
at backward angles. In other words, the suppression of the backward cross sections of the  $\Sn116(p,p)$ and $\Sn118(p,p)$ reactions is the signal of an enhanced $3s_{1/2}$ neutron component, indicating the 
the shell effect at $N=66$ in the Sn isotopes.
The neutron rms radii and skin thicknesses obtained in  the present analysis
show a smooth $N$ dependence along the isotope chain, and they are approximately consistent with the 
theoretical predictions of the RHB calculation with the me2 interaction. 
This seems to contradict the experimental results of Ref.~\cite{Terashima:2008zza}. 

In the experimental Sn$(p,p)$ data at 295~MeV in Ref.~\cite{Terashima:2008zza}, 
the normalization of the $\Sn120(p,p)$ and $\Sn124(p,p)$ cross section data was found to be inconsistent with 
the data for other isotopes, suggesting that they should be corrected. 
To extract further accurate values of the neutron skin thickness from the Sn$(p,p)$ data, 
a reanalysis taking into account the isotopic systematics of the data observed at 295~MeV 
is needed.

\begin{acknowledgments}
This work was supported
by Grants-in-Aid of the Japan Society for the Promotion of Science (Grant Nos. JP18K03617 and 18H05407)
and by a grant of the joint research project of the Research Center for Nuclear Physics at Osaka
University.
\end{acknowledgments}

\bibliographystyle{apsrev4-1}
\bibliography{RIA-pb-refs}

\begin{thebibliography}{32}%
\makeatletter
\providecommand \@ifxundefined [1]{%
 \@ifx{#1\undefined}
}%
\providecommand \@ifnum [1]{%
 \ifnum #1\expandafter \@firstoftwo
 \else \expandafter \@secondoftwo
 \fi
}%
\providecommand \@ifx [1]{%
 \ifx #1\expandafter \@firstoftwo
 \else \expandafter \@secondoftwo
 \fi
}%
\providecommand \natexlab [1]{#1}%
\providecommand \enquote  [1]{``#1''}%
\providecommand \bibnamefont  [1]{#1}%
\providecommand \bibfnamefont [1]{#1}%
\providecommand \citenamefont [1]{#1}%
\providecommand \href@noop [0]{\@secondoftwo}%
\providecommand \href [0]{\begingroup \@sanitize@url \@href}%
\providecommand \@href[1]{\@@startlink{#1}\@@href}%
\providecommand \@@href[1]{\endgroup#1\@@endlink}%
\providecommand \@sanitize@url [0]{\catcode `\\12\catcode `\$12\catcode
  `\&12\catcode `\#12\catcode `\^12\catcode `\_12\catcode `\%12\relax}%
\providecommand \@@startlink[1]{}%
\providecommand \@@endlink[0]{}%
\providecommand \url  [0]{\begingroup\@sanitize@url \@url }%
\providecommand \@url [1]{\endgroup\@href {#1}{\urlprefix }}%
\providecommand \urlprefix  [0]{URL }%
\providecommand \Eprint [0]{\href }%
\providecommand \doibase [0]{http://dx.doi.org/}%
\providecommand \selectlanguage [0]{\@gobble}%
\providecommand \bibinfo  [0]{\@secondoftwo}%
\providecommand \bibfield  [0]{\@secondoftwo}%
\providecommand \translation [1]{[#1]}%
\providecommand \BibitemOpen [0]{}%
\providecommand \bibitemStop [0]{}%
\providecommand \bibitemNoStop [0]{.\EOS\space}%
\providecommand \EOS [0]{\spacefactor3000\relax}%
\providecommand \BibitemShut  [1]{\csname bibitem#1\endcsname}%
\let\auto@bib@innerbib\@empty
\bibitem [{\citenamefont {Roca-Maza}\ \emph {et~al.}(2011)\citenamefont
  {Roca-Maza}, \citenamefont {Centelles}, \citenamefont {Vinas},\ and\
  \citenamefont {Warda}}]{RocaMaza:2011pm}%
  \BibitemOpen
  \bibfield  {author} {\bibinfo {author} {\bibfnamefont {X.}~\bibnamefont
  {Roca-Maza}}, \bibinfo {author} {\bibfnamefont {M.}~\bibnamefont
  {Centelles}}, \bibinfo {author} {\bibfnamefont {X.}~\bibnamefont {Vinas}}, \
  and\ \bibinfo {author} {\bibfnamefont {M.}~\bibnamefont {Warda}},\ }\href
  {\doibase 10.1103/PhysRevLett.106.252501} {\bibfield  {journal} {\bibinfo
  {journal} {Phys. Rev. Lett.}\ }\textbf {\bibinfo {volume} {106}},\ \bibinfo
  {pages} {252501} (\bibinfo {year} {2011})},\ \Eprint
  {http://arxiv.org/abs/1103.1762} {arXiv:1103.1762 [nucl-th]} \BibitemShut
  {NoStop}%
\bibitem [{\citenamefont {Roca-Maza}\ and\ \citenamefont
  {Paar}(2018)}]{Roca-Maza:2018ujj}%
  \BibitemOpen
  \bibfield  {author} {\bibinfo {author} {\bibfnamefont {X.}~\bibnamefont
  {Roca-Maza}}\ and\ \bibinfo {author} {\bibfnamefont {N.}~\bibnamefont
  {Paar}},\ }\href {\doibase 10.1016/j.ppnp.2018.04.001} {\bibfield  {journal}
  {\bibinfo  {journal} {Prog. Part. Nucl. Phys.}\ }\textbf {\bibinfo {volume}
  {101}},\ \bibinfo {pages} {96} (\bibinfo {year} {2018})},\ \Eprint
  {http://arxiv.org/abs/1804.06256} {arXiv:1804.06256 [nucl-th]} \BibitemShut
  {NoStop}%
\bibitem [{\citenamefont {Tsang}\ \emph {et~al.}(2012)\citenamefont {Tsang}
  \emph {et~al.}}]{Tsang:2012se}%
  \BibitemOpen
  \bibfield  {author} {\bibinfo {author} {\bibfnamefont {M.~B.}\ \bibnamefont
  {Tsang}} \emph {et~al.},\ }\href {\doibase 10.1103/PhysRevC.86.015803}
  {\bibfield  {journal} {\bibinfo  {journal} {Phys. Rev. C}\ }\textbf {\bibinfo
  {volume} {86}},\ \bibinfo {pages} {015803} (\bibinfo {year} {2012})},\
  \Eprint {http://arxiv.org/abs/1204.0466} {arXiv:1204.0466 [nucl-ex]}
  \BibitemShut {NoStop}%
\bibitem [{\citenamefont {Ray}(1979)}]{Ray:1979qv}%
  \BibitemOpen
  \bibfield  {author} {\bibinfo {author} {\bibfnamefont {L.}~\bibnamefont
  {Ray}},\ }\href {\doibase 10.1103/PhysRevC.20.1212} {\bibfield  {journal}
  {\bibinfo  {journal} {Phys. Rev. C}\ }\textbf {\bibinfo {volume} {19}},\
  \bibinfo {pages} {1855} (\bibinfo {year} {1979})},\ \bibinfo {note}
  {[Erratum: Phys.Rev.C 20, 1212--1212 (1979)]}\BibitemShut {NoStop}%
\bibitem [{\citenamefont {Terashima}\ \emph {et~al.}(2008)\citenamefont
  {Terashima} \emph {et~al.}}]{Terashima:2008zza}%
  \BibitemOpen
  \bibfield  {author} {\bibinfo {author} {\bibfnamefont {S.}~\bibnamefont
  {Terashima}} \emph {et~al.},\ }\href {\doibase 10.1103/PhysRevC.77.024317}
  {\bibfield  {journal} {\bibinfo  {journal} {Phys. Rev. C}\ }\textbf {\bibinfo
  {volume} {77}},\ \bibinfo {pages} {024317} (\bibinfo {year}
  {2008})}\BibitemShut {NoStop}%
\bibitem [{\citenamefont {Starodubsky}\ and\ \citenamefont
  {Hintz}(1994)}]{Starodubsky:1994xt}%
  \BibitemOpen
  \bibfield  {author} {\bibinfo {author} {\bibfnamefont {V.~E.}\ \bibnamefont
  {Starodubsky}}\ and\ \bibinfo {author} {\bibfnamefont {N.~M.}\ \bibnamefont
  {Hintz}},\ }\href {\doibase 10.1103/PhysRevC.49.2118} {\bibfield  {journal}
  {\bibinfo  {journal} {Phys. Rev. C}\ }\textbf {\bibinfo {volume} {49}},\
  \bibinfo {pages} {2118} (\bibinfo {year} {1994})}\BibitemShut {NoStop}%
\bibitem [{\citenamefont {Zenihiro}\ \emph {et~al.}(2010)\citenamefont
  {Zenihiro} \emph {et~al.}}]{Zenihiro:2010zz}%
  \BibitemOpen
  \bibfield  {author} {\bibinfo {author} {\bibfnamefont {J.}~\bibnamefont
  {Zenihiro}} \emph {et~al.},\ }\href {\doibase 10.1103/PhysRevC.82.044611}
  {\bibfield  {journal} {\bibinfo  {journal} {Phys. Rev. C}\ }\textbf {\bibinfo
  {volume} {82}},\ \bibinfo {pages} {044611} (\bibinfo {year}
  {2010})}\BibitemShut {NoStop}%
\bibitem [{\citenamefont {Trzcinska}\ \emph {et~al.}(2001)\citenamefont
  {Trzcinska}, \citenamefont {Jastrzebski}, \citenamefont {Lubinski},
  \citenamefont {Hartmann}, \citenamefont {Schmidt}, \citenamefont {von
  Egidy},\ and\ \citenamefont {Klos}}]{Trzcinska:2001sy}%
  \BibitemOpen
  \bibfield  {author} {\bibinfo {author} {\bibfnamefont {A.}~\bibnamefont
  {Trzcinska}}, \bibinfo {author} {\bibfnamefont {J.}~\bibnamefont
  {Jastrzebski}}, \bibinfo {author} {\bibfnamefont {P.}~\bibnamefont
  {Lubinski}}, \bibinfo {author} {\bibfnamefont {F.~J.}\ \bibnamefont
  {Hartmann}}, \bibinfo {author} {\bibfnamefont {R.}~\bibnamefont {Schmidt}},
  \bibinfo {author} {\bibfnamefont {T.}~\bibnamefont {von Egidy}}, \ and\
  \bibinfo {author} {\bibfnamefont {B.}~\bibnamefont {Klos}},\ }\href {\doibase
  10.1103/PhysRevLett.87.082501} {\bibfield  {journal} {\bibinfo  {journal}
  {Phys. Rev. Lett.}\ }\textbf {\bibinfo {volume} {87}},\ \bibinfo {pages}
  {082501} (\bibinfo {year} {2001})}\BibitemShut {NoStop}%
\bibitem [{\citenamefont {Klos}\ \emph {et~al.}(2007)\citenamefont {Klos} \emph
  {et~al.}}]{Klos:2007is}%
  \BibitemOpen
  \bibfield  {author} {\bibinfo {author} {\bibfnamefont {B.}~\bibnamefont
  {Klos}} \emph {et~al.},\ }\href {\doibase 10.1103/PhysRevC.76.014311}
  {\bibfield  {journal} {\bibinfo  {journal} {Phys. Rev. C}\ }\textbf {\bibinfo
  {volume} {76}},\ \bibinfo {pages} {014311} (\bibinfo {year} {2007})},\
  \Eprint {http://arxiv.org/abs/nucl-ex/0702016} {arXiv:nucl-ex/0702016}
  \BibitemShut {NoStop}%
\bibitem [{\citenamefont {Abrahamyan}\ \emph {et~al.}(2012)\citenamefont
  {Abrahamyan} \emph {et~al.}}]{Abrahamyan:2012gp}%
  \BibitemOpen
  \bibfield  {author} {\bibinfo {author} {\bibfnamefont {S.}~\bibnamefont
  {Abrahamyan}} \emph {et~al.},\ }\href {\doibase
  10.1103/PhysRevLett.108.112502} {\bibfield  {journal} {\bibinfo  {journal}
  {Phys. Rev. Lett.}\ }\textbf {\bibinfo {volume} {108}},\ \bibinfo {pages}
  {112502} (\bibinfo {year} {2012})},\ \Eprint {http://arxiv.org/abs/1201.2568}
  {arXiv:1201.2568 [nucl-ex]} \BibitemShut {NoStop}%
\bibitem [{\citenamefont {Friedman}(2012)}]{Friedman:2012pa}%
  \BibitemOpen
  \bibfield  {author} {\bibinfo {author} {\bibfnamefont {E.}~\bibnamefont
  {Friedman}},\ }\href {\doibase 10.1016/j.nuclphysa.2012.09.007} {\bibfield
  {journal} {\bibinfo  {journal} {Nucl. Phys. A}\ }\textbf {\bibinfo {volume}
  {896}},\ \bibinfo {pages} {46} (\bibinfo {year} {2012})},\ \Eprint
  {http://arxiv.org/abs/1209.6168} {arXiv:1209.6168 [nucl-ex]} \BibitemShut
  {NoStop}%
\bibitem [{\citenamefont {Tarbert}\ \emph {et~al.}(2014)\citenamefont {Tarbert}
  \emph {et~al.}}]{Tarbert:2013jze}%
  \BibitemOpen
  \bibfield  {author} {\bibinfo {author} {\bibfnamefont {C.~M.}\ \bibnamefont
  {Tarbert}} \emph {et~al.},\ }\href {\doibase 10.1103/PhysRevLett.112.242502}
  {\bibfield  {journal} {\bibinfo  {journal} {Phys. Rev. Lett.}\ }\textbf
  {\bibinfo {volume} {112}},\ \bibinfo {pages} {242502} (\bibinfo {year}
  {2014})},\ \Eprint {http://arxiv.org/abs/1311.0168} {arXiv:1311.0168
  [nucl-ex]} \BibitemShut {NoStop}%
\bibitem [{\citenamefont {Krasznahorkay}\ \emph {et~al.}(1999)\citenamefont
  {Krasznahorkay} \emph {et~al.}}]{Krasznahorkay:1999zz}%
  \BibitemOpen
  \bibfield  {author} {\bibinfo {author} {\bibfnamefont {A.}~\bibnamefont
  {Krasznahorkay}} \emph {et~al.},\ }\href {\doibase
  10.1103/PhysRevLett.82.3216} {\bibfield  {journal} {\bibinfo  {journal}
  {Phys. Rev. Lett.}\ }\textbf {\bibinfo {volume} {82}},\ \bibinfo {pages}
  {3216} (\bibinfo {year} {1999})}\BibitemShut {NoStop}%
\bibitem [{\citenamefont {Tamii}\ \emph {et~al.}(2011)\citenamefont {Tamii}
  \emph {et~al.}}]{Tamii:2011pv}%
  \BibitemOpen
  \bibfield  {author} {\bibinfo {author} {\bibfnamefont {A.}~\bibnamefont
  {Tamii}} \emph {et~al.},\ }\href {\doibase 10.1103/PhysRevLett.107.062502}
  {\bibfield  {journal} {\bibinfo  {journal} {Phys. Rev. Lett.}\ }\textbf
  {\bibinfo {volume} {107}},\ \bibinfo {pages} {062502} (\bibinfo {year}
  {2011})},\ \Eprint {http://arxiv.org/abs/1104.5431} {arXiv:1104.5431
  [nucl-ex]} \BibitemShut {NoStop}%
\bibitem [{\citenamefont {Piekarewicz}\ \emph {et~al.}(2012)\citenamefont
  {Piekarewicz}, \citenamefont {Agrawal}, \citenamefont {Colo}, \citenamefont
  {Nazarewicz}, \citenamefont {Paar}, \citenamefont {Reinhard}, \citenamefont
  {Roca-Maza},\ and\ \citenamefont {Vretenar}}]{Piekarewicz:2012pp}%
  \BibitemOpen
  \bibfield  {author} {\bibinfo {author} {\bibfnamefont {J.}~\bibnamefont
  {Piekarewicz}}, \bibinfo {author} {\bibfnamefont {B.~K.}\ \bibnamefont
  {Agrawal}}, \bibinfo {author} {\bibfnamefont {G.}~\bibnamefont {Colo}},
  \bibinfo {author} {\bibfnamefont {W.}~\bibnamefont {Nazarewicz}}, \bibinfo
  {author} {\bibfnamefont {N.}~\bibnamefont {Paar}}, \bibinfo {author}
  {\bibfnamefont {P.~G.}\ \bibnamefont {Reinhard}}, \bibinfo {author}
  {\bibfnamefont {X.}~\bibnamefont {Roca-Maza}}, \ and\ \bibinfo {author}
  {\bibfnamefont {D.}~\bibnamefont {Vretenar}},\ }\href {\doibase
  10.1103/PhysRevC.85.041302} {\bibfield  {journal} {\bibinfo  {journal} {Phys.
  Rev. C}\ }\textbf {\bibinfo {volume} {85}},\ \bibinfo {pages} {041302}
  (\bibinfo {year} {2012})},\ \Eprint {http://arxiv.org/abs/1201.3807}
  {arXiv:1201.3807 [nucl-th]} \BibitemShut {NoStop}%
\bibitem [{\citenamefont {Ray}\ \emph {et~al.}(1978)\citenamefont {Ray},
  \citenamefont {Coker},\ and\ \citenamefont {Hoffmann}}]{Ray:1978ws}%
  \BibitemOpen
  \bibfield  {author} {\bibinfo {author} {\bibfnamefont {L.}~\bibnamefont
  {Ray}}, \bibinfo {author} {\bibfnamefont {W.~R.}\ \bibnamefont {Coker}}, \
  and\ \bibinfo {author} {\bibfnamefont {G.~W.}\ \bibnamefont {Hoffmann}},\
  }\href {\doibase 10.1103/PhysRevC.18.2641} {\bibfield  {journal} {\bibinfo
  {journal} {Phys. Rev. C}\ }\textbf {\bibinfo {volume} {18}},\ \bibinfo
  {pages} {2641} (\bibinfo {year} {1978})}\BibitemShut {NoStop}%
\bibitem [{\citenamefont {Hoffmann}\ \emph {et~al.}(1980)\citenamefont
  {Hoffmann} \emph {et~al.}}]{Hoffmann:1980kg}%
  \BibitemOpen
  \bibfield  {author} {\bibinfo {author} {\bibfnamefont {G.~W.}\ \bibnamefont
  {Hoffmann}} \emph {et~al.},\ }\href {\doibase 10.1103/PhysRevC.21.1488}
  {\bibfield  {journal} {\bibinfo  {journal} {Phys. Rev. C}\ }\textbf {\bibinfo
  {volume} {21}},\ \bibinfo {pages} {1488} (\bibinfo {year}
  {1980})}\BibitemShut {NoStop}%
\bibitem [{\citenamefont {Zenihiro}\ \emph {et~al.}(2018)\citenamefont
  {Zenihiro} \emph {et~al.}}]{Zenihiro:2018rmz}%
  \BibitemOpen
  \bibfield  {author} {\bibinfo {author} {\bibfnamefont {J.}~\bibnamefont
  {Zenihiro}} \emph {et~al.},\ }\href@noop {} {\  (\bibinfo {year} {2018})},\
  \Eprint {http://arxiv.org/abs/1810.11796} {arXiv:1810.11796 [nucl-ex]}
  \BibitemShut {NoStop}%
\bibitem [{\citenamefont {Horowitz}(1985)}]{Horowitz:1985tw}%
  \BibitemOpen
  \bibfield  {author} {\bibinfo {author} {\bibfnamefont {C.~J.}\ \bibnamefont
  {Horowitz}},\ }\href {\doibase 10.1103/PhysRevC.31.1340} {\bibfield
  {journal} {\bibinfo  {journal} {Phys. Rev. C}\ }\textbf {\bibinfo {volume}
  {31}},\ \bibinfo {pages} {1340} (\bibinfo {year} {1985})}\BibitemShut
  {NoStop}%
\bibitem [{\citenamefont {Murdock}\ and\ \citenamefont
  {Horowitz}(1987)}]{Murdock:1986fs}%
  \BibitemOpen
  \bibfield  {author} {\bibinfo {author} {\bibfnamefont {D.~P.}\ \bibnamefont
  {Murdock}}\ and\ \bibinfo {author} {\bibfnamefont {C.~J.}\ \bibnamefont
  {Horowitz}},\ }\href {\doibase 10.1103/PhysRevC.35.1442} {\bibfield
  {journal} {\bibinfo  {journal} {Phys. Rev. C}\ }\textbf {\bibinfo {volume}
  {35}},\ \bibinfo {pages} {1442} (\bibinfo {year} {1987})}\BibitemShut
  {NoStop}%
\bibitem [{\citenamefont {Horowitz}\ \emph {et~al.}(1991)\citenamefont
  {Horowitz}, \citenamefont {Murdock},\ and\ \citenamefont
  {B.D.}}]{RIAcode:1991}%
  \BibitemOpen
  \bibfield  {author} {\bibinfo {author} {\bibfnamefont {C.}~\bibnamefont
  {Horowitz}}, \bibinfo {author} {\bibfnamefont {D.}~\bibnamefont {Murdock}}, \
  and\ \bibinfo {author} {\bibfnamefont {S.}~\bibnamefont {B.D.}},\ }\href@noop
  {} {\emph {\bibinfo {title} {Computational Nuclear Physics 1}}},\ edited by\
  \bibinfo {editor} {\bibfnamefont {K.}~\bibnamefont {Langanke}}, \bibinfo
  {editor} {\bibfnamefont {J.}~\bibnamefont {Maruhn}}, \ and\ \bibinfo {editor}
  {\bibfnamefont {S.}~\bibnamefont {Koonin}}\ (\bibinfo  {publisher}
  {Springer-Verlag},\ \bibinfo {year} {1991})\ p.\ \bibinfo {pages}
  {129}\BibitemShut {NoStop}%
\bibitem [{\citenamefont {Sakaguchi}\ \emph {et~al.}(1998)\citenamefont
  {Sakaguchi} \emph {et~al.}}]{Sakaguchi:1998zz}%
  \BibitemOpen
  \bibfield  {author} {\bibinfo {author} {\bibfnamefont {H.}~\bibnamefont
  {Sakaguchi}} \emph {et~al.},\ }\href {\doibase 10.1103/PhysRevC.57.1749}
  {\bibfield  {journal} {\bibinfo  {journal} {Phys. Rev. C}\ }\textbf {\bibinfo
  {volume} {57}},\ \bibinfo {pages} {1749} (\bibinfo {year}
  {1998})}\BibitemShut {NoStop}%
\bibitem [{\citenamefont {Kanada-En'yo}(2021)}]{Kanada-Enyo:2021}%
  \BibitemOpen
  \bibfield  {author} {\bibinfo {author} {\bibfnamefont {Y.}~\bibnamefont
  {Kanada-En'yo}},\ }\href@noop {} {\  (\bibinfo {year} {2021})},\ \Eprint
  {http://arxiv.org/abs/2106.00151} {arXiv:2106.00151 [nucl-th]} \BibitemShut
  {NoStop}%
\bibitem [{\citenamefont {Nik\v{s}i\'{c}}\ \emph {et~al.}(2014)\citenamefont
  {Nik\v{s}i\'{c}}, \citenamefont {Paar}, \citenamefont {Vretenar},\ and\
  \citenamefont {Ring}}]{Niksic:2014dra}%
  \BibitemOpen
  \bibfield  {author} {\bibinfo {author} {\bibfnamefont {T.}~\bibnamefont
  {Nik\v{s}i\'{c}}}, \bibinfo {author} {\bibfnamefont {N.}~\bibnamefont
  {Paar}}, \bibinfo {author} {\bibfnamefont {D.}~\bibnamefont {Vretenar}}, \
  and\ \bibinfo {author} {\bibfnamefont {P.}~\bibnamefont {Ring}},\ }\href
  {\doibase 10.1016/j.cpc.2014.02.027} {\bibfield  {journal} {\bibinfo
  {journal} {Comput. Phys. Commun.}\ }\textbf {\bibinfo {volume} {185}},\
  \bibinfo {pages} {1808} (\bibinfo {year} {2014})},\ \Eprint
  {http://arxiv.org/abs/1403.4039} {arXiv:1403.4039 [nucl-th]} \BibitemShut
  {NoStop}%
\bibitem [{\citenamefont {Bennaceur}\ and\ \citenamefont
  {Dobaczewski}(2005)}]{Bennaceur:2005mx}%
  \BibitemOpen
  \bibfield  {author} {\bibinfo {author} {\bibfnamefont {K.}~\bibnamefont
  {Bennaceur}}\ and\ \bibinfo {author} {\bibfnamefont {J.}~\bibnamefont
  {Dobaczewski}},\ }\href {\doibase 10.1016/j.cpc.2005.02.002} {\bibfield
  {journal} {\bibinfo  {journal} {Comput. Phys. Commun.}\ }\textbf {\bibinfo
  {volume} {168}},\ \bibinfo {pages} {96} (\bibinfo {year} {2005})},\ \Eprint
  {http://arxiv.org/abs/nucl-th/0501002} {arXiv:nucl-th/0501002} \BibitemShut
  {NoStop}%
\bibitem [{\citenamefont {Lalazissis}\ \emph {et~al.}(2005)\citenamefont
  {Lalazissis}, \citenamefont {Nik\v{s}i\'{c}}, \citenamefont {Vretenar},\ and\
  \citenamefont {Ring}}]{Lalazissis:2005de}%
  \BibitemOpen
  \bibfield  {author} {\bibinfo {author} {\bibfnamefont {G.~A.}\ \bibnamefont
  {Lalazissis}}, \bibinfo {author} {\bibfnamefont {T.}~\bibnamefont
  {Nik\v{s}i\'{c}}}, \bibinfo {author} {\bibfnamefont {D.}~\bibnamefont
  {Vretenar}}, \ and\ \bibinfo {author} {\bibfnamefont {P.}~\bibnamefont
  {Ring}},\ }\href {\doibase 10.1103/PhysRevC.71.024312} {\bibfield  {journal}
  {\bibinfo  {journal} {Phys. Rev. C}\ }\textbf {\bibinfo {volume} {71}},\
  \bibinfo {pages} {024312} (\bibinfo {year} {2005})}\BibitemShut {NoStop}%
\bibitem [{\citenamefont {Nik\v{s}i\'{c}}\ \emph {et~al.}(2008)\citenamefont
  {Nik\v{s}i\'{c}}, \citenamefont {Vretenar},\ and\ \citenamefont
  {Ring}}]{Niksic:2008vp}%
  \BibitemOpen
  \bibfield  {author} {\bibinfo {author} {\bibfnamefont {T.}~\bibnamefont
  {Nik\v{s}i\'{c}}}, \bibinfo {author} {\bibfnamefont {D.}~\bibnamefont
  {Vretenar}}, \ and\ \bibinfo {author} {\bibfnamefont {P.}~\bibnamefont
  {Ring}},\ }\href {\doibase 10.1103/PhysRevC.78.034318} {\bibfield  {journal}
  {\bibinfo  {journal} {Phys. Rev. C}\ }\textbf {\bibinfo {volume} {78}},\
  \bibinfo {pages} {034318} (\bibinfo {year} {2008})},\ \Eprint
  {http://arxiv.org/abs/0809.1375} {arXiv:0809.1375 [nucl-th]} \BibitemShut
  {NoStop}%
\bibitem [{\citenamefont {Bartel}\ \emph {et~al.}(1982)\citenamefont {Bartel},
  \citenamefont {Quentin}, \citenamefont {Brack}, \citenamefont {Guet},\ and\
  \citenamefont {Hakansson}}]{Bartel:1982ed}%
  \BibitemOpen
  \bibfield  {author} {\bibinfo {author} {\bibfnamefont {J.}~\bibnamefont
  {Bartel}}, \bibinfo {author} {\bibfnamefont {P.}~\bibnamefont {Quentin}},
  \bibinfo {author} {\bibfnamefont {M.}~\bibnamefont {Brack}}, \bibinfo
  {author} {\bibfnamefont {C.}~\bibnamefont {Guet}}, \ and\ \bibinfo {author}
  {\bibfnamefont {H.~B.}\ \bibnamefont {Hakansson}},\ }\href {\doibase
  10.1016/0375-9474(82)90403-1} {\bibfield  {journal} {\bibinfo  {journal}
  {Nucl. Phys. A}\ }\textbf {\bibinfo {volume} {386}},\ \bibinfo {pages} {79}
  (\bibinfo {year} {1982})}\BibitemShut {NoStop}%
\bibitem [{\citenamefont {Chabanat}\ \emph {et~al.}(1998)\citenamefont
  {Chabanat}, \citenamefont {Bonche}, \citenamefont {Haensel}, \citenamefont
  {Meyer},\ and\ \citenamefont {Schaeffer}}]{Chabanat:1997un}%
  \BibitemOpen
  \bibfield  {author} {\bibinfo {author} {\bibfnamefont {E.}~\bibnamefont
  {Chabanat}}, \bibinfo {author} {\bibfnamefont {P.}~\bibnamefont {Bonche}},
  \bibinfo {author} {\bibfnamefont {P.}~\bibnamefont {Haensel}}, \bibinfo
  {author} {\bibfnamefont {J.}~\bibnamefont {Meyer}}, \ and\ \bibinfo {author}
  {\bibfnamefont {R.}~\bibnamefont {Schaeffer}},\ }\href {\doibase
  10.1016/S0375-9474(98)00180-8} {\bibfield  {journal} {\bibinfo  {journal}
  {Nucl. Phys. A}\ }\textbf {\bibinfo {volume} {635}},\ \bibinfo {pages} {231}
  (\bibinfo {year} {1998})},\ \bibinfo {note} {[Erratum: Nucl.Phys.A 643,
  441--441 (1998)]}\BibitemShut {NoStop}%
\bibitem [{\citenamefont {Angeli}\ and\ \citenamefont
  {Marinova}(2013)}]{Angeli:2013epw}%
  \BibitemOpen
  \bibfield  {author} {\bibinfo {author} {\bibfnamefont {I.}~\bibnamefont
  {Angeli}}\ and\ \bibinfo {author} {\bibfnamefont {K.~P.}\ \bibnamefont
  {Marinova}},\ }\href {\doibase 10.1016/j.adt.2011.12.006} {\bibfield
  {journal} {\bibinfo  {journal} {Atom. Data Nucl. Data Tabl.}\ }\textbf
  {\bibinfo {volume} {99}},\ \bibinfo {pages} {69} (\bibinfo {year}
  {2013})}\BibitemShut {NoStop}%
\bibitem [{\citenamefont {El~Bassem}\ and\ \citenamefont
  {Oulne}(2019)}]{Bassem:2019gil}%
  \BibitemOpen
  \bibfield  {author} {\bibinfo {author} {\bibfnamefont {Y.}~\bibnamefont
  {El~Bassem}}\ and\ \bibinfo {author} {\bibfnamefont {M.}~\bibnamefont
  {Oulne}},\ }\href {\doibase 10.1016/j.nuclphysa.2019.04.003} {\bibfield
  {journal} {\bibinfo  {journal} {Nucl. Phys. A}\ }\textbf {\bibinfo {volume}
  {987}},\ \bibinfo {pages} {16} (\bibinfo {year} {2019})},\ \Eprint
  {http://arxiv.org/abs/1904.10318} {arXiv:1904.10318 [nucl-th]} \BibitemShut
  {NoStop}%
\bibitem [{\citenamefont {Haider}\ \emph {et~al.}(2010)\citenamefont {Haider},
  \citenamefont {Sharma}, \citenamefont {Gambhir},\ and\ \citenamefont
  {Kailas}}]{Haider:2010zz}%
  \BibitemOpen
  \bibfield  {author} {\bibinfo {author} {\bibfnamefont {W.}~\bibnamefont
  {Haider}}, \bibinfo {author} {\bibfnamefont {M.}~\bibnamefont {Sharma}},
  \bibinfo {author} {\bibfnamefont {Y.~K.}\ \bibnamefont {Gambhir}}, \ and\
  \bibinfo {author} {\bibfnamefont {S.}~\bibnamefont {Kailas}},\ }\href
  {\doibase 10.1103/PhysRevC.81.034601} {\bibfield  {journal} {\bibinfo
  {journal} {Phys. Rev. C}\ }\textbf {\bibinfo {volume} {81}},\ \bibinfo
  {pages} {034601} (\bibinfo {year} {2010})}\BibitemShut {NoStop}%
\end{thebibliography}%

\end{document}